Manuscript

# Autonomous scanning electrochemical cell microscopy enables rapid exploration of large compositionally complex material spaces

Felix Thelen[1‡], Moonjoo Kim[2‡], Geovane Arruda de Oliveira[2], Jan Lukas Bürgel[1], Wolfgang Schuhmann[2,*], Alfred Ludwig[1,*]

[1]Chair for Materials Discovery and Interfaces, Institute for Materials, Faculty of Mechanical Engineering, Ruhr University Bochum, Universitätsstraße 150, 44801 Bochum, Germany

[2]Analytical Chemistry - Center for Electrochemical Sciences (CES), Faculty of Chemistry and Biochemistry, Ruhr University Bochum, Universitätsstraße 150, 44801 Bochum, Germany

[‡]These authors contributed equally.

*corresponding authors: alfred.ludwig@rub.de; wolfgang.schuhmann@rub.de

## Abstract

Alloying is a central strategy in electrocatalysis, enabling fine-tuning of electronic structure. In particular, compositionally complex solid solutions (CCSS) often called high-entropy alloys are of high interest as they allow active site design. However, the "combinatorial explosion" in the number of possible compositions poses a critical bottleneck for the discovery of active CCSS electrocatalysts. We present an autonomous scanning electrochemical cell microscopy (SECCM) system for ultrahigh-throughput and large-scale CCSS activity screening. The platform rapidly establishes composition–electrocatalytic activity relationships for large compositional spaces across multiple thin-film CCSS materials libraries via active learning and automated library exchange. Embedding analytical expressions of voltammetry in the algorithm enables the learning of whole voltammograms rather than a single selected metric. As a demonstration, we investigated hydrogen evolution reaction (HER) activities of Au–Ir–Rh, where Ir and Rh exhibit strong metal-hydrogen binding and Au exhibits relatively weak binding as derived from the HER volcano plot. The composition-activity trend was accurately predicted after measuring only 15% of all 966 measurement areas. $Au_{30}Ir_{20}Rh_{50}$ and $Au_{10}Ir_{35}Rh_{55}$ exhibit highest activities with standard rate constants of ~0.012 cm $s^{-1}$, demonstrating positive synergistic contributions from elemental mixing. The autonomous robotic SECCM platform is broadly applicable to a wide range of electrocatalytic reactions, providing a general pathway for accelerating CCSS electrocatalyst discovery and optimization.

## 1. Introduction

Compositionally complex solid solutions (CCSS) allow for finely tuning the electronic structure of active catalytic sites, thereby enabling quasi-continuous adjustments in the binding strength of adsorbates [1]. This tunability positions CCSS as a promising platform for developing high-activity electrocatalysts. However, the enormous design space for CCSS electrocatalysts poses a fundamental challenge for experimental exploration [2,3]. When considering a resolution of 5 at.%, a quaternary system already contains 1771 possible compositions, while a quinary system expands to 10.626 compositions.

Combinatorial thin-film synthesis techniques, such as magnetron sputtering, can address this challenge by enabling rapid fabrication (within hours) of thousands of CCSS electrocatalysts in the form of well-defined compositional gradients, so-called thin-film materials libraries. By pairing this synthesis with high-throughput characterization methods that allow automated measurement of hundreds of compositions during several hours, this combinatorial approach significantly improves efficiency compared to serial, one-at-a-time experiments [4,5]. Energy-dispersive X-ray spectroscopy (EDX) for compositional analysis, X-ray diffraction (XRD) for phase identification, and functional property characterization can all be completed in just over two days per materials library containing 342 measurement areas, which each refer to an investigated composition [6].

Thin-film CCSS materials libraries provide a well-controlled platform for assessing composition-dependent intrinsic electrocatalytic activity. The uniform morphology on a library minimizes extrinsic contributions which could arise from variations in particle size and shape that are commonly encountered in nanoparticle systems. The planar geometry also simplifies interfacial mass transport, facilitating the accurate evaluation of electrocatalytic reaction kinetics. Combining materials libraries with high-throughput electrochemical characterization techniques such as scanning droplet cell (SDC) and long-range scanning electrochemical cell microscopy (SECCM) has brought substantial progress in CCSS electrocatalyst discovery. The SDC enabled automated electrochemical screening of materials libraries on timescales of days [7]. Long-range (e.g., across 100 mm diameter materials libraries) SECCM advanced this approach by using nanoscale capillaries rather than a droplet cell geometry, resulting in improved data quality, industrially relevant current densities, and the possibility of subsequent characterization due to its minimal surface footprint, all while maintaining a comparable measurement throughput [8]. This integrated multimodal high-throughput characterization workflow has accelerated the discovery of high-activity CCSS electrocatalysts and helped investigating composition-structure-activity relationships of CCSS [9,10].

Despite these advances, systemically exploring the full compositional landscape of CCSS electrocatalysts is still exceptionally challenging. A single materials library fabricated by magnetron co-sputtering in the center of a ternary composition space covers approximately 20% of the composition space, a single library in a quaternary composition space 5%, while a single library in the quinary space covers less than 1% [6]. Consequently, identifying new electrocatalysts in compositionally complex materials requires the fabrication and characterization of numerous libraries.

Recent advances in algorithm-guided experimental design have demonstrated that Bayesian optimization and active learning can dramatically reduce the number of experiments required to navigate high-dimensional composition spaces. In electrocatalysis, experimental Bayesian optimization has been applied to identify active compositions in compositionally complex materials, including both thin-film libraries and nanoparticle catalysts [11,12]. Beyond electrocatalysis, autonomous experimentation has advanced rapidly across the broader chemical sciences. Mobile robotic platforms and self-driving laboratories in combination with Bayesian optimization allow autonomous experiments [13–16]. However, most approaches operate on sequentially synthesized individual samples, limiting their scalability compared to combinatorial thin-film libraries. Furthermore, electrochemical performance is typically assessed through single metrics extracted at fixed potentials, discarding the wealth of kinetic information encoded in the full voltammetric response.

Here, we present an autonomous robotic long-range SECCM platform that can rapidly learn composition-dependent electrocatalytic activities of CCSS systems. A robot arm-assisted automated exchange of the materials libraries is integrated with long-range SECCM: Instead of sequentially screening the electrocatalytic activities of each prepared materials library, the new system can seamlessly traverse multiple materials libraries in a single operation. The Gaussian process-based active learning algorithm guides the platform in dynamically selecting and measuring the electrocatalytic behavior of an informative CCSS composition, reducing the number of data points required to determine the composition–activity trend of the system. We integrated multi-output Gaussian process regression with the analytical expression of voltammetry for the SECCM measurements so that the platform can “learn” the linear sweep voltammogram (LSV) for HER and quantitatively extract kinetics descriptors from the measurements. Correcting for the mass transport factor allows for a more accurate assessment of electrocatalytic activities than relying on single-metric evaluations such as current density. As a demonstration, we used the ternary system Au–Ir–Ru as a representative model. Its elements were chosen based on the hypothesis that synergy may arise from pairing metals with stronger M–H binding strength than the optimum value (Ir and Rh) with a metal exhibiting weaker M–H binding strength (Au).

## 2. Results and Discussion

### 2.1. Design of the robotic SECCM for autonomous electrochemical screening

The design of the long-range robotic SECCM platform is shown in Figure 1. The platform consists of a long-range SECCM [8] for performing electrochemical measurements across the materials libraries (Figure 1a) and a cassette for storing the CCSS materials libraries (Figure 1b). A six-axis robotic arm enables the automated transfer of materials libraries between the storage cassette and the SECCM, allowing automated measurements across of (currently) up to ten libraries without human intervention. The materials libraries are mounted on custom-designed sample holders to be securely handled by the robotic arm. To ensure accurate positioning and electrical contact, the sample holders and table are designed with chamfered edges automatically aligning the sample holders upon placement of the robot. Gold-plated pins and fixation screws ensure electrical contact to ground. The entire system is placed inside a grounded Faraday cage to minimize noise, i.e. increase the signal to noise ratio. The SECCM capillary, which contains an electrolyte and a reference-counter electrode, forms a sub-microscale electrochemical cell. The tip of the capillary is located inside a 3D-printed environmental chamber which provides a continuous humidified Ar-atmosphere around the capillary tip, which allows investigating HER without interference from the oxygen reduction reaction (ORR). The environmental chamber is automatically positioned above the sample based on the feedback of a force sensor and remains at a distance of 0.1 mm to the sample’s surface at all times. The capillary is approached to the surface in an automated way as well. By detecting and observing the capillary tip and its reflection in the live feed of a microscopy camera with computer vision techniques, the capillary is positioned 100 μm above the sample. The fine approach to the nanometer level is subsequently performed with a piezo actuator. A detailed description about the instrument’s components is provided in Section 1.6 of the Supporting Information (Figures S1–S5)

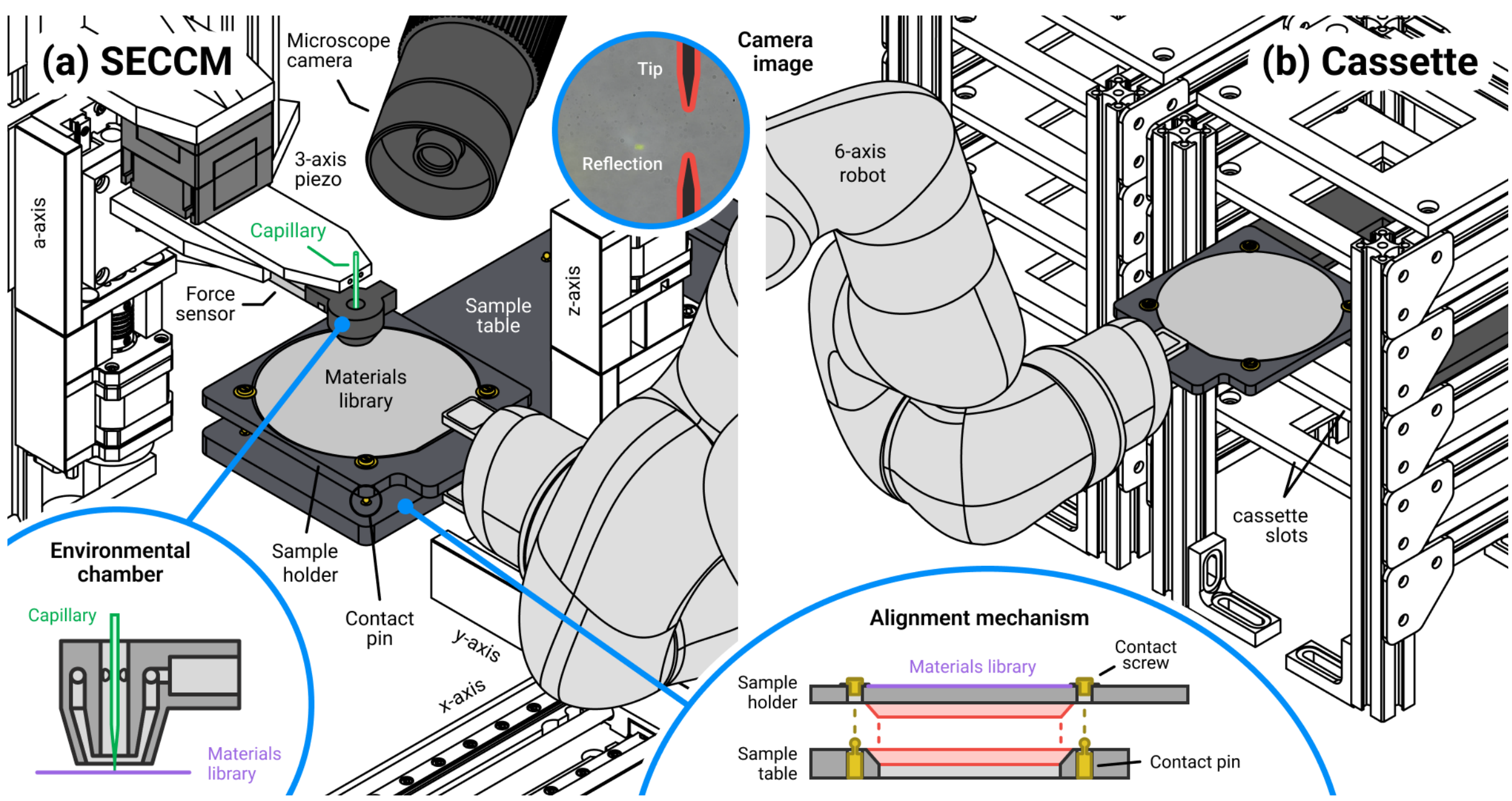

**Figure 1:** Components of the robotic SECCM. The device consists of an SECCM illustrated in (a), and the cassette setup for 10 libraries in (b), connected through a six-axis robot to enable cross-library screening. The SECCM consists of a capillary surrounded by a 3D-printed environmental chamber enabling a localized humidified Ar-atmosphere around the tip. By taking the feedback of a force sensor attached to the chamber and a visual representation of the tip and its reflection through a microscope camera into account, the device performs an automated approach of the tip to the surface of the materials library. Upon placement of the robotic arm, an alignment and contact mechanism built into the sample table and holder enables accurate positioning. For a detailed description of the instrument setup and the control software see Section 1.6 of the Supporting Information and Figures S1–S5.

The workflow of active learning for autonomous SECCM measurements is summarized in Figure 2**.** Following the robotic arm's automated library loading, the system performs tilt correction and surface approach procedures before initiating electrochemical measurements. Each newly acquired linear sweep voltammogram (LSV) is fitted with an analytical equation based on Butler-Volmer kinetics under steady-state mass transport [17] to decouple the individual electrochemical kinetics descriptors, namely the standard rate constant ($k^0$) and the transfer coefficient ($\alpha$), from mass transfer effects. The current density reflects both electrochemical activity and mass transport effects [18]. Although often overlooked in high-throughput electrochemical screening, material-dependent variations in mass transport, arising from factors such as real surface area [19] and electrode morphology [20], should be corrected for accurate electrocatalytic activity evaluation. In SECCM, the droplet cell wetting area can depend on the compositional-dependent wettability of the investigated material, leading to variations in mass transport (Figure S6) [21]. These kinetics descriptors are used as output training data for a multi-output Gaussian process, functioning as the surrogate model in an active learning campaign. This allows learning and predicting the HER LSV shape of each measured material, instead of operating on single scalar values obtained from the LSVs like current densities at a fixed potential, which was the prior state of the art. The volume compositions of all measurement areas acquired by EDX are used as input training data. By combining predicted model uncertainty with estimated measurement time for each area, the algorithm autonomously selects the next measurement area and determines when to transition between libraries. After reaching a predefined stopping criterion or receiving user input to halt ("human-in-the-loop"), the system can predict LSVs for unmeasured compositions using the trained Gaussian process model and the analytical LSV equation. This approach is highly scalable, with 10 cassette slots currently implemented and straightforward expansion is possible through hardware additions.

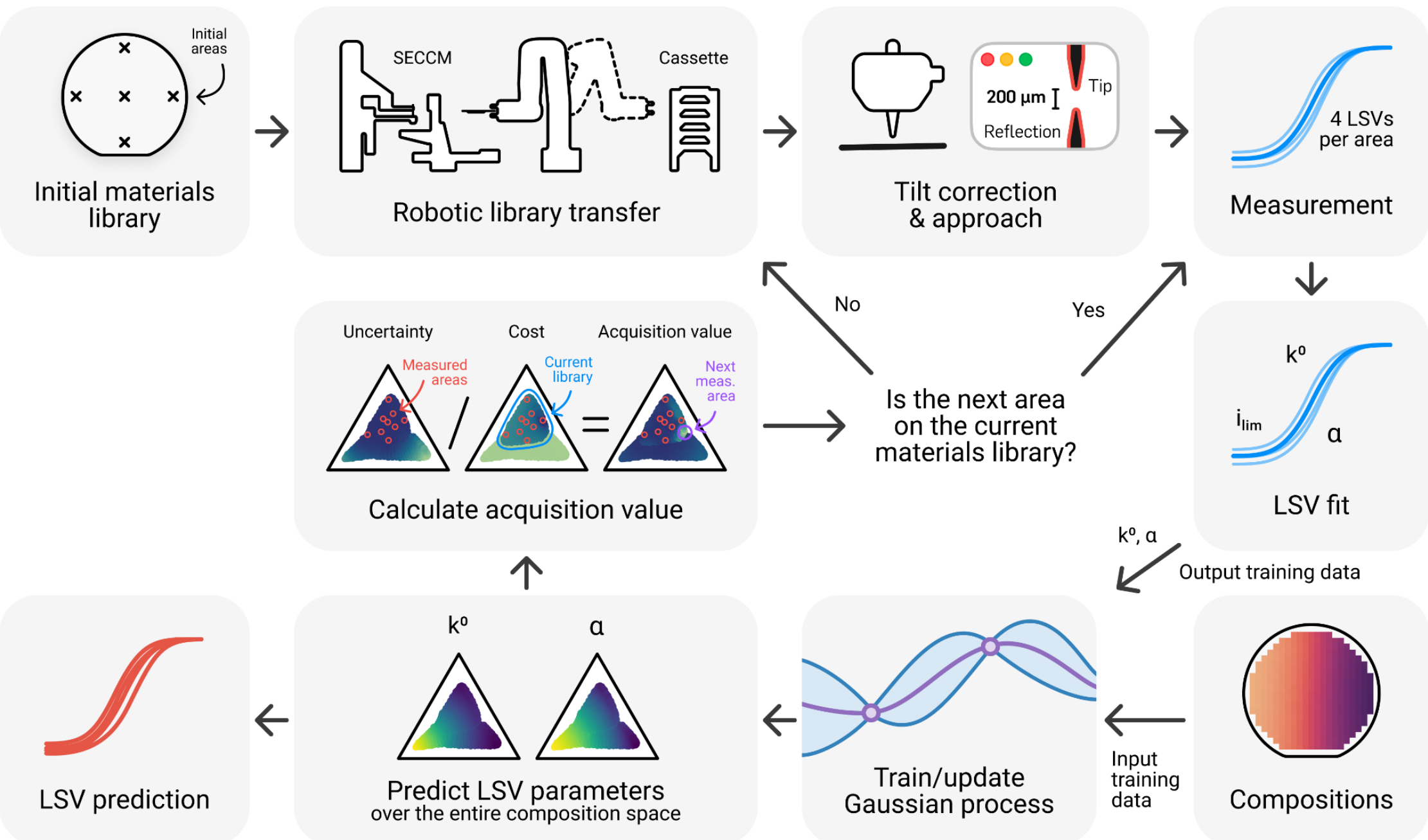

**Figure 2:** Measurement workflow of the autonomous SECCM platform. The system contains a robotic arm that autonomously transfers materials libraries between a storage cassette and the SECCM. Following automated tilt correction and surface approach, the platform records LSVs across a library. Each LSV is fitted with an analytical function to extract physically interpretable descriptors, which are used to train a multi-output Gaussian process model within a cost-aware active learning framework. The trained model predicts the electrochemical behavior for unmeasured compositions and autonomously guides exploration of the composition space by selecting subsequent measurement areas and determining optimal timing for library transitions.

### 2.2. Compositional and structural characterization of Au–Ir–Rh

Whereas the autonomous SECCM is applicable to materials libraries with arbitrary compositional complexity, we demonstrate the performance of the autonomous robotic SECCM on three materials libraries fabricated in the ternary composition space Au–Ir–Rh. This system includes constituent elements from both sides of the acidic HER volcano plot: Ir and Rh bind hydrogen intermediates too strongly, while Au binds them too weakly [22]. A ternary system was chosen over higher-order ones to simplify visualization and interpretation of composition–activity relationships. Three Au–Ir–Rh libraries were fabricated by magnetron co-sputtering covering 63% of the overall ternary composition space. Details of the thin-film synthesis are provided in the Supporting Information.

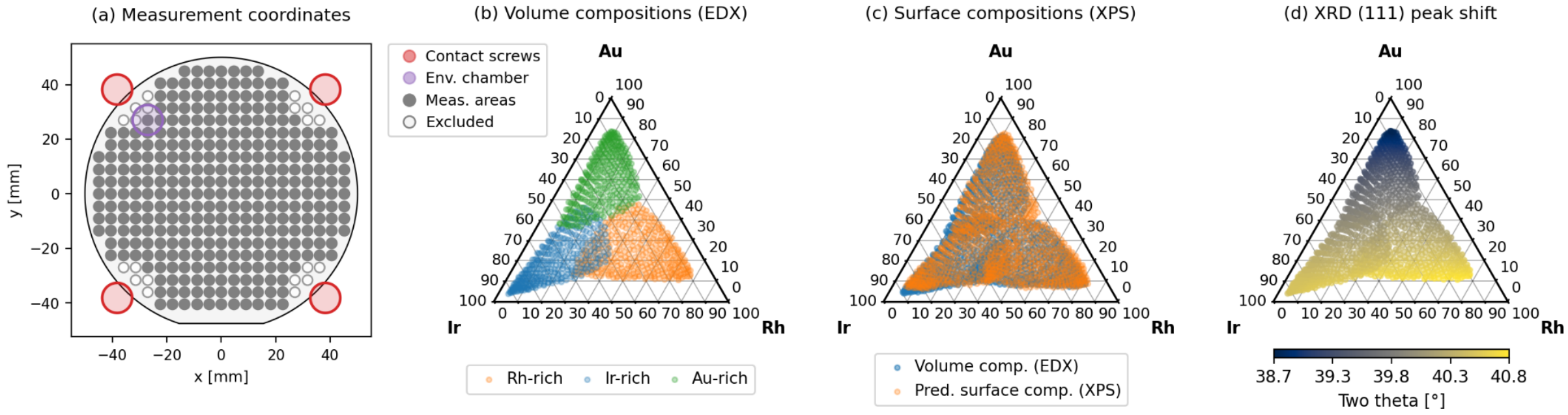

**Figure 3:** Characterization of the three Au–Ir–Rh thin-film materials libraries. (a) The 342-measurement area grid used for all characterization techniques. 20 areas are excluded from electrochemical measurement to avoid collisions between the environmental chamber and the sample fixation screws; (b) Composition spreads of the three libraries plotted in the ternary composition space acquired by EDX. (c) Predicted surface compositions based on 39 local XPS measurements on the three libraries. Surface and volume compositions are in close agreement, indicating that the EDX composition can be used as a proxy for the surface compositions in active learning driven SECCM measurements. (d) XRD peak shift of the most prominent (111) reflection over the composition space. The continuous peak shift across the libraries confirms the presence of a compositionally complex solid solution in all measurement areas. An extended analysis of the EDX, XPS and XRD results is provided in the Sections 2.1–2.3 in the Supporting Information.

A 342-measurement area grid was used for all characterization techniques as shown in Figure 3a. Due to the geometry of the sample mounting and the footprint of the environmental chamber, 20 measurement areas were excluded from electrochemical characterization to avoid collisions during lateral translation. Figure 3b shows the volume composition spreads of the three libraries determined by EDX projected onto the ternary composition space. Each library is enriched in one of the three elements, and — as part of our experimental strategy — partial compositional overlaps exist between the Ir-rich and Rh-rich libraries as well as between the Ir-rich and Au-rich libraries, providing regions for cross-validation of the observed property trends. As the probing depth of EDX extends into the micrometer range, the measured compositions represent volume rather than surface compositions. For each library, 13 complementary, evenly distributed XPS measurements were performed to assess the validity of using the XPS values as a proxy for the surface composition, as the electrocatalytic reactions take place at the surface. Using multi-output Gaussian process prediction [23], the 39 local XPS measurements were extended to the full measurement grid (1026 measurement areas), and the resulting surface compositions are shown together with the volume compositions in Figure 3c. The surface and volume compositions are in close agreement, with a mean Euclidean distance of 4.2 at.%. Additionally, native Rh-oxide and Rh-hydroxide peaks were identified in the XPS spectra at >30 at.% Rh-content, with a native Rh-oxide content of up to 20 at.% and a native Rh-hydroxide content of up to 8 at.%. A detailed analysis of the surface composition distributions is provided in the Section 2.2 of the Supporting Information. To confirm that the observed activity distributions can be attributed to compositional trends and not to possible variations of the phase constitution, X-ray diffraction (XRD) measurements were performed. XRD confirmed a single-phase solid solution across the full compositional range of all three libraries. Figure 3d shows the peak shift of the most prominent XRD reflection around 40° 2θ over the composition space, which was assigned to the (111) reflection of an fcc lattice. The peak shift smoothly extends over all three libraries in accordance with lattice parameter differences of elemental Au, Ir and Rh. In addition, the thin-film morphology was assessed with 15 localized atomic force microscopy (AFM) measurements spread across the explored composition space as a proxy for the electrochemical surface area. The extracted arithmetic mean roughness ranges from 0.7 to 1.6 nm, confirming that the films are smooth at the nanoscale and that absolute differences in surface area between the compositions are small. A detailed XRD and AFM analysis is provided in the Supporting Information Sections 2.3 and 2.4, respectively.

### 2.3. High-throughput measurement of HER activity in Au–Ir–Rh

Figure 4a shows the LSVs acquired during the autonomous measurement of all three libraries. In total 966 measurement areas were characterized without manual intervention, using a capillary with an opening diameter of 510 nm. The total measurement duration was ~26 h. Details of the measurement parameters are provided in the Section 1.6–1.7 of the Supporting Information. The LSVs show a key advantage of SECCM over other high-throughput electrochemical techniques: the small confinement of the electrochemical cell enables industrially relevant current densities of several A cm$^{-2}$, and the LSVs show a well-defined spherical diffusion-limited plateau current. Additionally, the low noise levels across all measurements are a prerequisite for the analytical fitting of the LSVs.

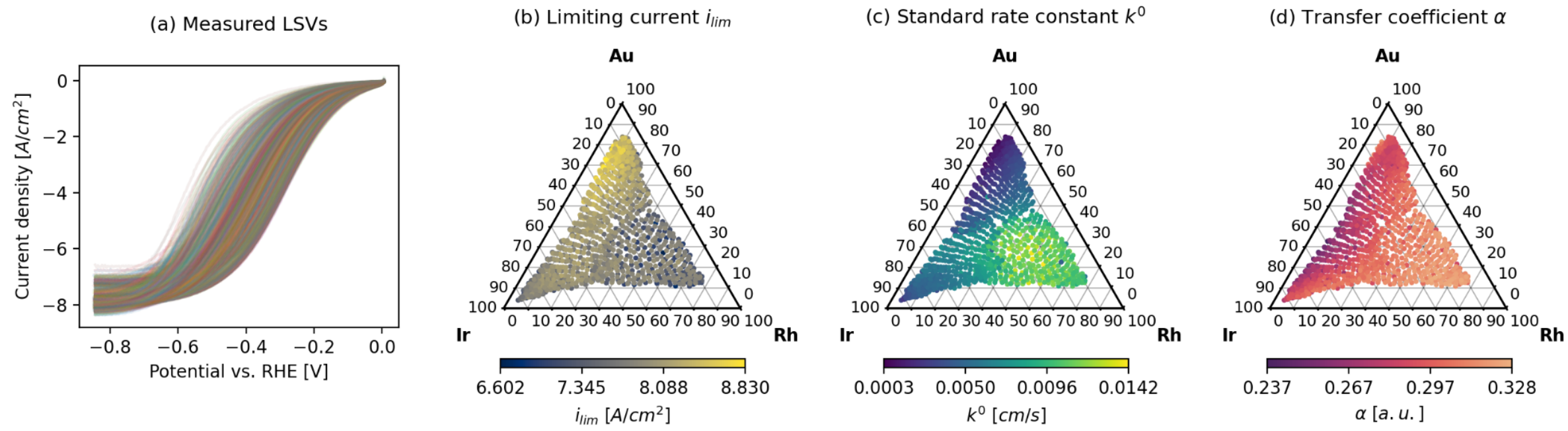

**Figure 4:** (a) Measured LSVs from all measurement areas of the three materials libraries. Distributions of the LSV-fitted parameters across the Au–Ir–Rh composition space: (b) limiting current density ($i_{lim}$), (c) standard rate constant $k^0$, and (d) transfer coefficient $\alpha$ for all 966 measurement areas across the three libraries. The smooth variation of all three parameters across library boundaries confirms consistency of the measurements. The limiting current density increases gradually with Au content and is the lowest at Rh-rich compositions. The standard rate constant shows a strongly non-linear compositional dependence, with a global maximum in the interior of the Rh-rich library that decreases toward all three binary edges and reaches its minimum at Au-rich compositions. The transfer coefficient varies more modestly, peaking at the highest Rh contents and decreasing toward Ir- and Au-rich regions.

The results of LSV fitting are shown in Figure 4, which projects the limiting current density $i_{lim}$ (b), $k^0$ (c), and $\alpha$ (d) onto the composition space. All three parameters vary smoothly and continuously both within and across library boundaries, with no discontinuities at the compositional overlaps between the Ir-rich, Rh-rich, and Au-rich libraries. This is related to the inherent nature of CCSS electrocatalysts, where the distribution of various surface atom arrangements or active sites in CCSS changes continuously as the composition varies [1].

The limiting current density, which is directly proportional to the mass transfer coefficient of the SECCM electrochemical cells, varies by up to 2.2 A cm$^{-2}$ across the full dataset, increases gradually with Au content and is lowest in Rh-rich regions. It shows local variations within individual libraries, most prominently in the Rh-rich library. Observing the limiting current over the number of measurements (Figure S21) reveals a dependence of the limiting current on library transfers. During each transfer, the capillary is briefly (4 min) removed from the controlled humidified Ar-atmosphere of the environmental chamber for repositioning and tilt correction. Changes of the electrolyte meniscus during this interval likely alter the effective wetted contact area before the next approach, thereby shifting the limiting current. In between transfers, the limiting current stays rather constant and only shifts minimally in dependence of the CCSS compositions. These measurement-order effects of the limiting current can also be observed in the distributions of $k^0$ and $\alpha$ (Figure 4b,c), however, to a much smaller extent. $k^0$ spans several orders of magnitude across the composition space, in line with preliminary experiments using other materials libraries. Its global maximum is located in the center of the Rh-rich library and decreases towards all three binary edges, reaching its lowest values at Au-rich compositions. This non-monotonic behavior suggests that the optimal kinetic performance for the HER in this system arises from synergistic alloying effects rather than from any single constituent element. In comparison, the transfer coefficient shows comparatively modest variation across the composition space: it is the highest at Rh-rich compositions and decreases gradually toward Ir- and Au-rich regions, with the minimum found in the Au-Ir-rich edge.

### 2.4. Noise-aware Gaussian process model for active learning of the HER activity

The active learning performance of the initial implementation was evaluated retrospectively against the ground truth of all 966 measurements. While the model captures the global compositional trends in both $k^0$ and $\alpha$ from early in the measurement campaign, an analysis reveals that the model overfits $k^0$ in the later stages of the campaign, with the normalized mean absolute error (MAE) approaching 0% while that of $\alpha$ stabilizes at ~5%. This overfitting is a consequence of the logarithmic transformation applied to $k^0$: by compressing nearly two orders of magnitude in linear space to a span of ~4 in log-space, the transformation reduces the replicate variance relative to the signal range, causing the Gaussian process noise parameter to converge to an artificially small value. A further contributing factor is the order in which averaging and log-transformation were applied — three LSVs were averaged per measurement area before log-transforming $k^0$, an operation which does not preserve the variance in log-space. As a result, the model misinterprets local fluctuations arising from the measurement noise and drift as genuine compositional features. The full analysis and evolution of the model predictions in this measurement campaign are provided in Section 2.6 in the Supporting Information.

To address the overfitting, the model was extended to incorporate the per-point measurement variance directly into the Gaussian likelihood. Rather than learning a single, shared noise parameter across all training points — as done in the first version of the algorithm — the updated likelihood accepts a fixed variance for each observation. This variance is obtained by fitting each of the three LSVs per measurement area separately with the analytical model, log-transforming the resulting $k^0$ values and computing the mean and variance in log-space. This ordering preserves the statistical meaning of the variance in the space in which the Gaussian process operates. As a result, areas with consistent measurement results constrain the predictions more tightly, while areas with higher noise are fitted more loosely.

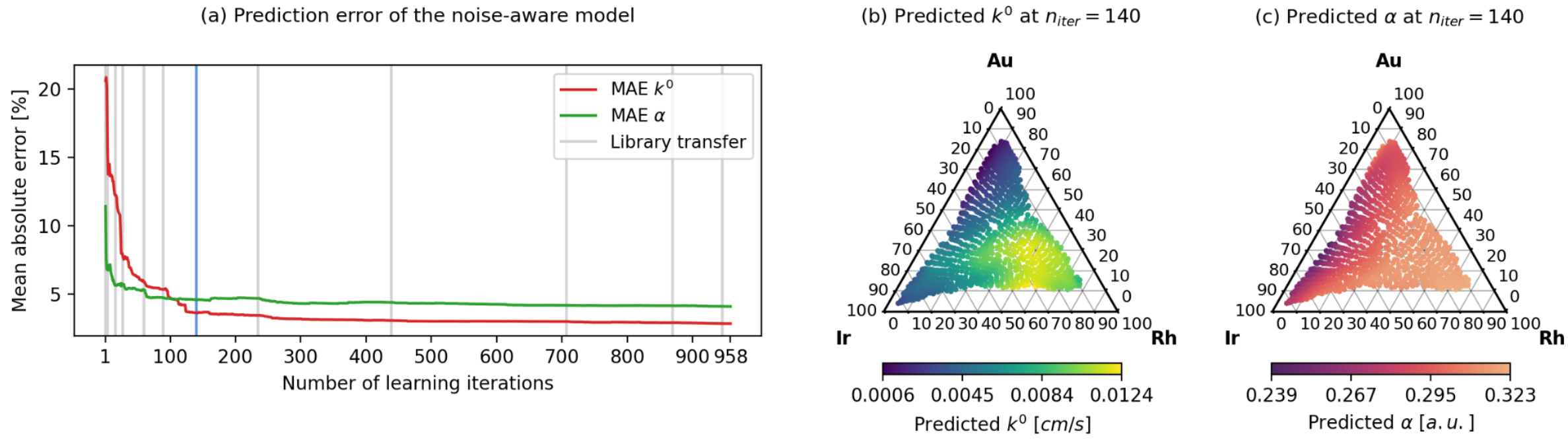

**Figure 6:** Active learning performance and predictions of the model with integrated per-point variance at iteration 140 (blue vertical line). (a) MAE prediction errors of $k^0$ and $\alpha$ for the noise aware model normalized to their specific ranges, (b) predicted $k^0$ distribution, (c) the predicted $\alpha$ distribution on the ternary composition space. Compared to the predictions of the original model (Figure S22d,e), the updated predictions show a smoother trend.

Since the modified likelihood alters the predictive variance and thereby the acquisition function, the active learning campaign was re-executed retrospectively on the existing dataset. Figure 6a shows the prediction MAE of the noise-aware model over the number of iterations. $k^0$ converges to a MAE of ~3%, and $\alpha$ to ~5%. For $k^0$, the largest improvement in predictive accuracy occurs within the first 100 iterations, where the MAE drops from 21% to 5%, after which the curve plateaus at 140 iterations and continues with only marginal further improvement. In contrast, the MAE of $\alpha$ starts lower at 13% and reaches its plateau around 5% after 60–100 iterations. Figure 6b,c show the predicted composition-activity maps for $k^0$ and $\alpha$ at iteration 140. Corresponding predictions at earlier and later iterations are provided in the Supporting Information Figure S23. Early iterations are dominated by prior uncertainty, while later iterations converge to visually indistinguishable maps. Compared to the predictions from the original model in Figure S22d,e, the updated predictions exhibit visibly smoother composition-dependent trends, most notably for $k^0$ and to a lesser extent for $\alpha$.

Judging from the MAE of $k^0$, the autonomous measurement could be stopped after approximately 100–140 iterations with minimal loss in accuracy. The MAE of $\alpha$ plateaus even earlier after 60 iterations. Since $k^0$ is the more relevant parameter for assessing electrocatalytic activity as it directly determines electrocatalytic performance, stopping after 140 iterations is justified, which reduces the number of required measurements to 15%.

In practice, during an autonomous SECCM experiment where ground truth is unavailable, the mean predictive variance can serve as a real-time stopping criterion, as demonstrated previously for autonomous electrical resistance measurements [24]. In the present dataset, the predictive variance plateaus at approximately the same iteration count at which the MAE of $k^0$ converges. Figure S24 shows the predictive variance over the number of iterations.

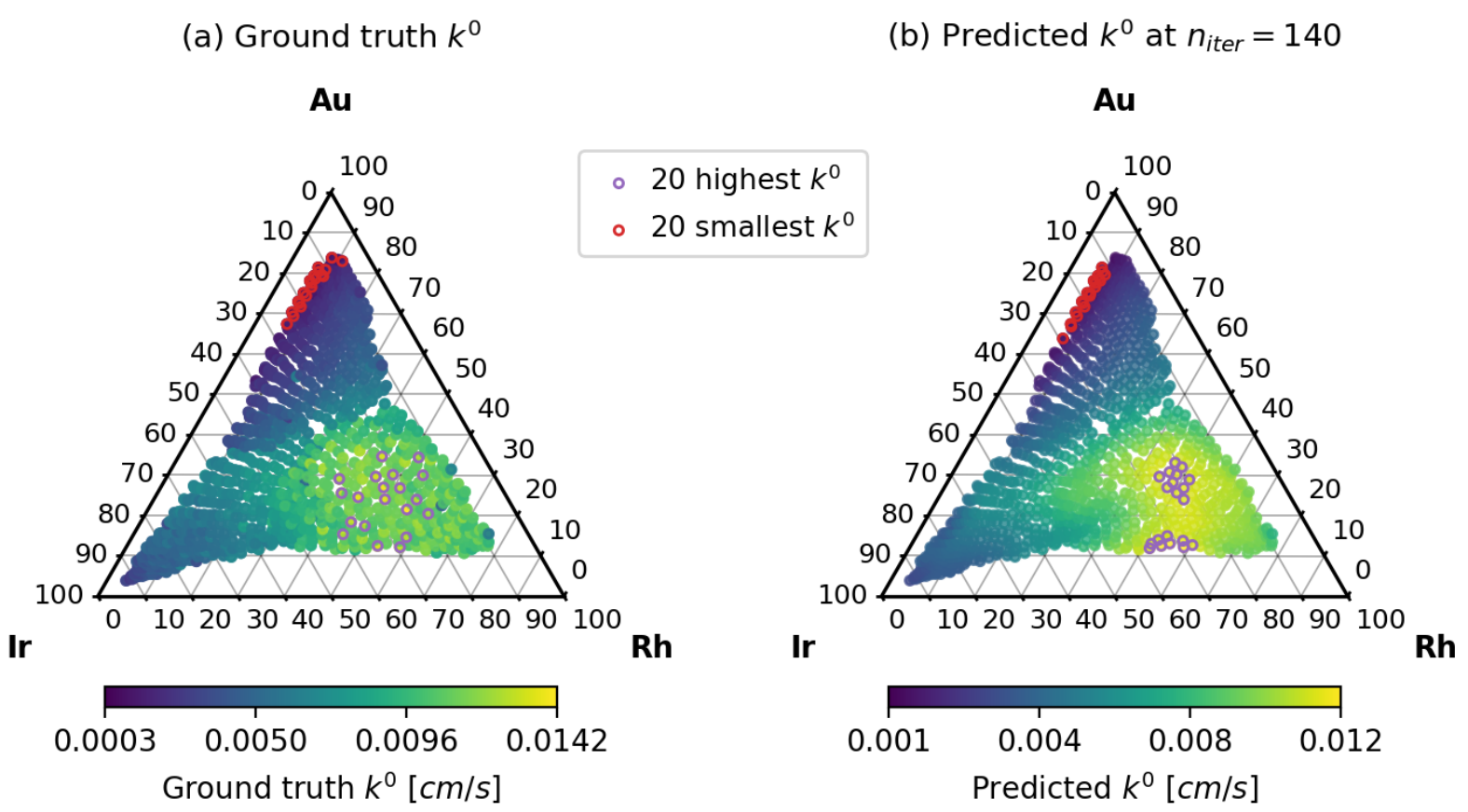

**Figure 7:** The 10 smallest (red) and highest (purple) standard rate constants $k^0$ in (a) the ground truth and (b) the predictions after 140 iterations.

Figure 7 compares the extracted $k^0$ values of the full ground truth dataset and the prediction of $k^0$ after 140 training iterations. The $Au_{60–80}Ir_{10–30}Rh_{10}$ composition range exhibits the lowest activity values in both Figure 7a and 7b. The highest activity compositions are located in the Rh-rich library. However, the maximum activity compositions of the raw measurement results are broadly dispersed (Figure 7a), corresponding to noise in the data. Through the predictions of the Gaussian process model, the trend (Figure 7b) reveals well-defined regions of high activity at compositions near $Au_{30}Ir_{20}Rh_{50}$ and $Au_{10}Ir_{35}Rh_{55}$ with $k^0$ values of approximately 0.012 cm $s^{-1}$. This corresponds to an exchange current density of 0.19 mA $cm^{-2}$ at pH 1 and 1.9 mA $cm^{-2}$ at pH 0. Although the activity is lower than that of Pt [25], these results support the initial design strategy of combining Au, Ir, and Rh, showing a synergistic effect that originates from mixing elements with strong and weak M–H interactions.

## 3. Conclusions

We have developed a fully autonomous robotic SECCM platform for high-throughput electrochemical characterization of combinatorial thin-film materials libraries and demonstrated its capabilities for accurately mapping large composition spaces on the example of three libraries spanning the ternary Au–Ir–Rh composition space. The platform integrates automated library exchange, tilt correction, and environmental atmosphere control into a single workflow, enabling unattended operation without manual intervention.

A key advantage of the approach over conventional high-throughput electrochemical screening methods is the use of the kinetic parameters, specifically the standard rate constant $k^0$ and the transfer coefficient $\alpha$ as figures of merit, rather than current densities extracted at a fixed overpotential. This approach is enabled by fitting the high-quality LSVs obtained by SECCM with an analytical equation and allows the underlying Gaussian process model behind the autonomous measurement protocol to predict LSVs from not yet measured areas based on the already acquired information.

The platform demonstrated high measurement robustness since all 966 measurement areas across the three libraries were characterized using a single nano-sized capillary. While the present demonstration used three of ten currently available library slots, the available space inside the Faraday cage would accommodate a substantially larger number of libraries (>50), offering a clear path towards further upscaling. The minimal footprint of the capillary droplet preserves the integrity of the thin-film surface and therefore enables subsequent characterization (e.g. for other reactions), which is not possible with other electrochemical characterization techniques.

The active learning framework further demonstrates that the number of measurements required to accurately infer the composition-activity landscape is a small fraction of the total library size. The noise-aware Gaussian process model is able to closely resemble the relative trends after ~50 iterations and achieves near-converged predictions of $k^0$ within 100–140 iterations, by on average 50 areas per library. Therefore, approximately 15% of all measurement areas are sufficient to reliably capture the composition-activity correlations across the Au–Ir–Rh composition space. On this basis, a single capillary rated at 1000 measurements could in principle support campaigns spanning up to 20 materials libraries in a single run.

While the present demonstration focused on HER, the platform is in principle applicable to a broad range of electrochemical reactions, provided that an appropriate analytical LSV model is established for parameter extraction. Taken together, the results establish the autonomous robotic SECCM as a scalable and resource-efficient platform for accelerated electrochemical materials discovery across complex composition spaces.

## Author contributions

F. Thelen and M. Kim contributed equally to this work. F. Thelen: conceptualization, methodology (measurement protocol, active learning framework, Gaussian process modeling, web-interface protocol), software (Python implementation, data analysis pipeline), formal analysis, data curation, validation, visualization (figures, animations, video production), investigation (SECCM measurements and analysis, EDX/XPS/XRD/AFM analysis), writing — original draft, writing — review and editing. M. Kim: conceptualization, methodology (electrochemical measurement protocol, web-interface protocol), software (LabView implementation), investigation (initial SECCM measurements and analysis), validation, writing — original draft, writing — review and editing. G. A. de Oliveira: investigation (SECCM measurements), writing — review and editing. J. L. Bürgel: investigation (sample fabrication, SECCM/EDX/XPS/XRD/AFM measurements, XPS/AFM analysis), writing — original draft, writing - review and editing. W. Schuhmann: conceptualization, supervision, resources, project administration, funding acquisition, writing - review and editing. A. Ludwig: conceptualization, supervision, resources, project administration, funding acquisition, writing - review and editing.

## Conflict of interest

There are no conflicts of interest to declare.

## Acknowledgements

This work was partially supported by different projects. A. Ludwig, F. Thelen and J. L. Bürgel acknowledge funding from the European Union through the European Research Council (ERC) Synergy Grant project 101118768 (DEMI). Views and opinions expressed are however those of the authors only and do not necessarily reflect those of the European Union or ERCEA. Neither the European Union nor the granting authority can be held responsible for them. Funding by the Deutsche Forschungsgemeinschaft (DFG, German Research

Foundation) CRC 1625, project number 506711657 is acknowledged by A. Ludwig, M. Kim, G. A. de Oliveira, and W. Schuhmann. The authors acknowledge using ZGH infrastructure for measurements (clean room (sputter deposition), electrochemistry lab, Scanning Electron Microscope, Atomic Force Microscope and X-Ray Diffractometer). The members of the mechanical workshop of the Faculty of Chemistry and Biochemistry, Ruhr University Bochum, are acknowledged for fabrication of the sample table and holders.

## Data availability

The data that support the findings of this study, including animations of both the original active learning model, the noise-aware one as well as a video of the measurement process are openly available in ZENODO at https://doi.org/10.5281/zenodo.20439519.

## References

[1] T. Löffler, A. Ludwig, J. Rossmeisl, W. Schuhmann, What Makes High-Entropy Alloys Exceptional Electrocatalysts?, Angew. Chem. Int. Ed. 60 (2021) 26894–26903. https://doi.org/10.1002/anie.202109212.

[2] L. Han, S. Zhu, Z. Rao, C. Scheu, D. Ponge, A. Ludwig, H. Zhang, O. Gutfleisch, H. Hahn, Z. Li, D. Raabe, Multifunctional High-Entropy Materials, Nat. Rev. Mater. 9 (2024) 846–865. https://doi.org/10.1038/s41578-024-00720-y.

[3] T.A.A. Batchelor, J.K. Pedersen, S.H. Winther, I.E. Castelli, K.W. Jacobsen, J. Rossmeisl, High-Entropy Alloys as a Discovery Platform for Electrocatalysis, Joule 3 (2019) 834–845. https://doi.org/10.1016/j.joule.2018.12.015.

[4] A. Ludwig, Discovery of New Materials Using Combinatorial Synthesis and High-Throughput Characterization of Thin-Film Materials Libraries Combined with Computational Methods, Npj Comput. Mater. 5 (2019) 70. https://doi.org/10.1038/s41524-019-0205-0.

[5] H. Koinuma, I. Takeuchi, Combinatorial Solid-State Chemistry of Inorganic Materials, Nature Mater. 3 (2004) 429–438. https://doi.org/10.1038/nmat1157.

[6] F. Thelen, R. Zehl, R. Zerdoumi, J.L. Bürgel, L. Banko, W. Schuhmann, A. Ludwig, Accelerating Combinatorial Electrocatalyst Discovery with Bayesian Optimization: a Case Study in the Quaternary System Ni-Pd-Pt-Ru for the Oxygen Evolution Reaction, Adv. Sci. 12 (2025) e07302. https://doi.org/10.1002/advs.202507302.

[7] K. Sliozberg, D. Schäfer, T. Erichsen, R. Meyer, C. Khare, A. Ludwig, W. Schuhmann, High-Throughput Screening of Thin-Film Semiconductor Material Libraries I: System Development and Case Study for Ti-W-O, ChemSusChem 8 (2015) 1270–1278. https://doi.org/10.1002/cssc.201402917.

[8] E.B. Tetteh, O.A. Krysiak, A. Savan, M. Kim, R. Zerdoumi, T.D. Chung, A. Ludwig, W. Schuhmann, Long-Range SECCM Enables High-Throughput Electrochemical Screening of High Entropy Alloy Electrocatalysts at Up-To-Industrial Current Densities, Small Meth. 8 (2024) 2301284. https://doi.org/10.1002/smtd.202301284.

[9] N. Pukhareva, M. Kim, F. Thelen, G. Arruda De Oliveira, R. Zehl, W. Schuhmann, A. Ludwig, Compositional and Structural Impact on the Hydrogen Evolution Reaction Activity across Noble-Metal-Based Compositionally Complex Solid Solutions Thin Film

Libraries, ACS Electrochem. 2 (2026) 364–373. https://doi.org/10.1021/acselectrochem.5c00406.

[10] J.L. Bürgel, R. Zehl, F. Thelen, R. Zerdoumi, O.A. Krysiak, B. Kohnen, E. Suhr, W. Schuhmann, A. Ludwig, Exploration of Nanostructured High-Entropy Alloys for Key Electrochemical Reactions: A Comparative Study for the Solid Solution Systems Cu–Pd–Pt–Ru, Ir–Pd–Pt–Ru and Ni–Pd–Pt–Ru, Faraday Discuss. 264 (2026) 64–82. https://doi.org/10.1039/D5FD00082C.

[11] V.A. Mints, J.K. Pedersen, A. Bagger, J. Quinson, A.S. Anker, K.M.Ø. Jensen, J. Rossmeisl, M. Arenz, Exploring the Composition Space of High-Entropy Alloy Nanoparticles for the Electrocatalytic $H_2$/CO Oxidation with Bayesian Optimization, ACS Catal. 12 (2022) 11263–11271. https://doi.org/10.1021/acscatal.2c02563.

[12] N. Namuersaihan, Z. Zhao, O.J. Conquest, Y. Shu, H. Sun, C. Su, Q. Cheng, A. Soon, C. Stampfl, J. Huang, Accelerated Discovery of High Performance $Ni_3S_4$/$Ni_3Mo$ HER Catalysts via Bayesian Optimization, Adv. Func. Mater. 36 (2026) e28363. https://doi.org/10.1002/adfm.202528363.

[13] T. Dai, S. Vijayakrishnan, F.T. Szczypiński, J.-F. Ayme, E. Simaei, T. Fellowes, R. Clowes, L. Kotopanov, C.E. Shields, Z. Zhou, J.W. Ward, A.I. Cooper, Autonomous Mobile Robots for Exploratory Synthetic Chemistry, Nature 635 (2024) 890–897. https://doi.org/10.1038/s41586-024-08173-7.

[14] B.P. MacLeod, F.G.L. Parlane, T.D. Morrissey, F. Häse, L.M. Roch, K.E. Dettelbach, R. Moreira, L.P.E. Yunker, M.B. Rooney, J.R. Deeth, V. Lai, G.J. Ng, H. Situ, R.H. Zhang, M.S. Elliott, T.H. Haley, D.J. Dvorak, A. Aspuru-Guzik, J.E. Hein, C.P. Berlinguette, Self-driving Laboratory for Accelerated Discovery of Thin-Film Materials, Sci. Adv. 6 (2020) eaaz8867. https://doi.org/10.1126/sciadv.aaz8867.

[15] Z. Zhang, Z. Ren, C.-W. Hsu, W. Chen, Z.-W. Hong, C.-F. Lee, A. Penn, H. Xu, D.J. Zheng, S. Miao, Y. Huang, Y. Gao, W. Chen, H. Smith, Y. Niu, Y. Tian, Y.-R. Lu, Y.-C. Shao, S. Li, H.-T. Wang, I.I. Abate, P. Agrawal, Y. Shao-Horn, J. Li, A Multimodal Robotic Platform for Multi-Element Electrocatalyst Discovery, Nature 647 (2025) 390–396. https://doi.org/10.1038/s41586-025-09640-5.

[16] B.P. MacLeod, F.G.L. Parlane, C.C. Rupnow, K.E. Dettelbach, M.S. Elliott, T.D. Morrissey, T.H. Haley, O. Proskurin, M.B. Rooney, N. Taherimakhsousi, D.J. Dvorak, H.N. Chiu, C.E.B. Waizenegger, K. Ocean, M. Mokhtari, C.P. Berlinguette, A Self-Driving Laboratory Advances the Pareto Front for Material Properties, Nat. Commun. 13 (2022) 995. https://doi.org/10.1038/s41467-022-28580-6.

[17] K.L. Anderson, M.A. Edwards, Evaluating Analytical Expressions for Scanning Electrochemical Cell Microscopy (SECCM), Anal. Chem. 95 (2023) 8258–8266. https://doi.org/10.1021/acs.analchem.3c00216.

[18] C. Wei, S. Sun, D. Mandler, X. Wang, S.Z. Qiao, Z.J. Xu, Approaches for Measuring the Surface Areas of Metal Oxide Electrocatalysts for Determining their Intrinsic Electrocatalytic Activity, Chem. Soc. Rev. 48 (2019) 2518–2534. https://doi.org/10.1039/C8CS00848E.

[19] C. Wei, R.R. Rao, J. Peng, B. Huang, I.E.L. Stephens, M. Risch, Z.J. Xu, Y. Shao-Horn, Recommended Practices and Benchmark Activity for Hydrogen and Oxygen Electrocatalysis in Water Splitting and Fuel Cells, Adv. Mater. 31 (2019) 1806296. https://doi.org/10.1002/adma.201806296.

[20] C. Lin, R.G. Compton, Understanding Mass Transport Influenced Electrocatalysis at the Nanoscale via Numerical Simulation, Cur. Opin. Electrochem. 14 (2019) 186–199. https://doi.org/10.1016/j.coelec.2018.08.001.

[21] G. Arruda De Oliveira, M. Kim, C.S. Santos, N. Limani, T.D. Chung, E.B. Tetteh, W. Schuhmann, Controlling Surface Wetting in High-Alkaline Electrolytes for Single Facet Pt Oxygen Evolution Electrocatalytic Activity Mapping by Scanning Electrochemical Cell Microscopy, Chem. Sci. 15 (2024) 16331–16337. https://doi.org/10.1039/D4SC04407J.

[22] S. Trasatti, Work function, Electronegativity, and Electrochemical Behaviour of Metals, J. Electroanal. Chem. Interfac. Electrochem. 39 (1972) 163–184. https://doi.org/10.1016/S0022-0728(72)80485-6.

[23] F. Thelen, F. Lourens, A. Ludwig, Accelerating Surface Composition Characterization of Thin-Film Materials Libraries Using Multi-Output Gaussian Process Regression, Adv. Intell. Discov. 2 (2026) e202500062. https://doi.org/10.1002/aidi.202500062.

[24] F. Thelen, L. Banko, R. Zehl, S. Baha, A. Ludwig, Speeding Up High-Throughput Characterization of Materials Libraries by Active Learning: Autonomous Electrical Resistance Measurements, Digit. Discov. 2 (2023) 1612–1619. https://doi.org/10.1039/D3DD00125C.

[25] E.B. Tetteh, M. Kim, A. Savan, A. Ludwig, T.D. Chung, W. Schuhmann, Reassessing the Intrinsic Hydrogen Evolution Reaction Activity of Platinum Using Scanning Electrochemical Cell Microscopy, Cell Rep. Phys. Sci. 4 (2023) 101680. https://doi.org/10.1016/j.xcrp.2023.101680.

Supporting Information

# Autonomous scanning electrochemical cell microscopy enables rapid exploration of large compositionally complex material spaces

Felix Thelen[1‡], Moonjoo Kim[2‡], Geovane Arruda de Oliveira[2], Jan Lukas Bürgel[1], Wolfgang Schuhmann[2*], Alfred Ludwig[1,*]

[1]Chair for Materials Discovery and Interfaces, Institute for Materials, Faculty of Mechanical Engineering, Ruhr University Bochum, Universitätsstraße 150, 44801 Bochum, Germany

[2]Analytical Chemistry - Center for Electrochemical Sciences (CES), Faculty of Chemistry and Biochemistry, Ruhr University Bochum, Universitätsstraße 150, 44801 Bochum, Germany

‡These authors contributed equally.

*corresponding authors: alfred.ludwig@rub.de; wolfgang.schuhmann@rub.de

## Table of contents

## 1. Experimental procedures

### 1.1. Synthesis of thin-film materials libraries synthesis in the system Au–Ir–Rh

The depositions of thin-film materials libraries were performed in a sputter system (“ATC2200”, AJA International, Inc.) with 8 cathodes (“A320-XP UHV”, AJA International). For Rh and Ir, pulsed DC power supplies (“DCXP 1500”, AJA International) were used in DC-mode. Au was sputtered using an RF-power supply (“0313 GTC”, connected to a “AIT-600 RF Auto Tuner”, both AJA International). The depositions were performed in Ar-atmosphere at a pressure of 0.667 Pa with a flow rate of 30 sccm. Elemental targets with a diameter of 50.8 mm were used: Au (Testbourne, purity: 99.99%, factory thickness: 3 mm), Rh (Tesbourne, purity: 99.95%, factory thickness: 2 mm), Ir (Alfa Aesar, purity: 99.9%, factory thickness: 3 mm), Ta (Testbourne, purity: 99.99%, factory thickness: 6.25 mm). Due to limited adhesion of noble metals on the sapphire substrate, a Ta adhesion layer (nominal thickness 10 nm) was deposited on the substrates (Si-Mat - Silicon Materials e.K., diameter: 100 mm, c-cut) prior to the deposition of the noble metal layer. The deposition parameters of each library and cathode can be found in Table S1. The deposition time was selected to obtain an average film thickness of 100 nm for each library.

**Table S1:** Deposition parameters for the three materials libraries. The corresponding research data management identification numbers for the libraries are 0011389 for the Au-rich library, 0011387 for the Ir-rich one and 0011388 for the Rh-rich library.

| | **Au** (cathode 1) RF power supply | | | **Ir** (cathode 6) DC power supply | | | **Rh** (cathode 3) DC power supply | | | |
|---|---|---|---|---|---|---|---|---|---|---|
| Library | Power [W] | Voltage [V] | Tune/ Load [%] | Power [W] | Voltage [V] | Current [A] | Power [W] | Voltage [V] | Current [A] | Time [s] |
| **Au-rich** | 83 | 63 | 56/13 | 37 | 296 | 0.13 | 36 | 246 | 0.153 | 1120 |
| **Ir-rich** | 29 | 45 | 54/22 | 80 | 324 | 0.25 | 26 | 247 | 0.112 | 1500 |
| **Rh-rich** | 23 | 42 | 54/21 | 20 | 288 | 0.075 | 55 | 260 | 0.218 | 2200 |

### 1.2. High-throughput energy dispersive X-ray spectroscopy (EDX)

The investigation of chemical composition was performed using a scanning electron microscope (SEM, “VEGA3”, Tescan) equipped with W filament and an EDX-detector (“XFlash 7”, Bruker). Each measurement area was analyzed with an acceleration voltage of 20 kV, a view field of 300 µm and a working distance of 15 mm. A deadtime of 25% was maintained throughout the screening. The acquired spectra were quantified using the “Esprit 2.6” Software (Bruker).

### 1.3. X-ray photoelectron spectroscopy (XPS)

XPS-measurements were conducted using an automated XPS (“Axis Nova”, Krathos) equipped with a monochromatic Al Kα radiation source operating at 180 W (15 mA emission current, 12 kV anode voltage) and a delay-line detector with 20 eV pass energy. High-resolution spectra were acquired for the Ir 4f, Au 4f and Rh 3d as well as the C 1s and the O 1s region. For quantification, a Shirley background subtraction was used alongside the systems built in relative sensitivity factors (“ESCApe”, Kratos). Charge referencing was done using the C 1s peak (adventitious C-C at 284.8 eV).

### 1.4. High-throughput X-ray diffraction (XRD)

The measurements were performed using a diffractometer (D8 Discover, Bruker), which is equipped with a Cu microfocus X-ray source (Incotec IµS High Brilliance, $\lambda$ = 0.15418 nm) and a 2D-Detector (Vantec-500, Bruker). The detector covers an angular range of 40°. Frames of diffraction data were collected at the $2\theta$ angles of 30°, 55°, and 80°, providing a combined coverage of 10° to 100°. The three recorded diffractogram frames were stitched together utilizing the “DIFFRAC.EVA” software (Bruker) and converted to a 1D-diffractogram. The software was further employed to subtract the background (threshold: 0.7, curvature: 0.3).

### 1.5. Atomic force microscopy (AFM)

For the roughness measurements, an AFM (“Fast Scan”, Bruker) was used with a “ScanAsyst-Air”-probe (Bruker, spring constant approximately 0.4 N $m^{-1}$) in the PeakForce Tapping mode with ScanAsyst. The software “NanoScope Analysis 1.8” (Bruker) was used to process the data and calculate the roughness of the thin films.

### 1.6. Autonomous scanning electrochemical cell microscopy (SECCM) platform

The robotic SECCM platform was developed based on the long-range SECCM reported in our previous work [1]. A single-barrel, laser-pulled quartz capillary with a tip-opening diameter of approximately 500 nm is positioned at the center of the instrument. The capillary is mechanically coupled to a three-axis piezo actuator (PiezoConcept C3.200), enabling translation in x, y, and z with a nanoscale positioning precision of 0.2 nm and a range of motion of 200 µm per axis. The capillary and piezo assembly are mounted on a rigid stand to kinematically decouple them from the remaining moving components of the system. The capillary is enclosed by a 3D-printed environmental chamber with an 8 mm bottom aperture, continuously purged with humidified Ar at a flow rate of 20 ml $min^{-1}$. The environmental chamber is mounted on a dedicated stepper motor, allowing it to be raised and lowered independently of the other axes. A Wheatstone bridge-type force sensor connected to the chamber enables reliable detection of surface contact upon approach.

The internal geometry of the environmental chamber is shown in Figure S1a. The capillary is inserted from above into a central tube of 4 mm inner diameter. A ring of radial channels connects to this central tube near the top of the chamber, directing a portion of the Ar flow upward through the gap between the capillary and the inner wall, thereby preventing ambient air from entering the chamber from above. The remaining Ar flow exits through the 8 mm bottom aperture, where a concentric ring of channels surrounds the capillary tip, maintaining an inert atmosphere at the droplet at the tip of the capillary. During measurements, the chamber is positioned 0.1 mm above the sample surface, which is small enough to suppress diffusion of ambient air into the chamber while avoiding mechanical contact with the sample.

(a) Environmental chamber design

Capillary

4 mm

Ar-flow

0.1 mm

Sample surface

8 mm

(b) Chamber approach algorithm

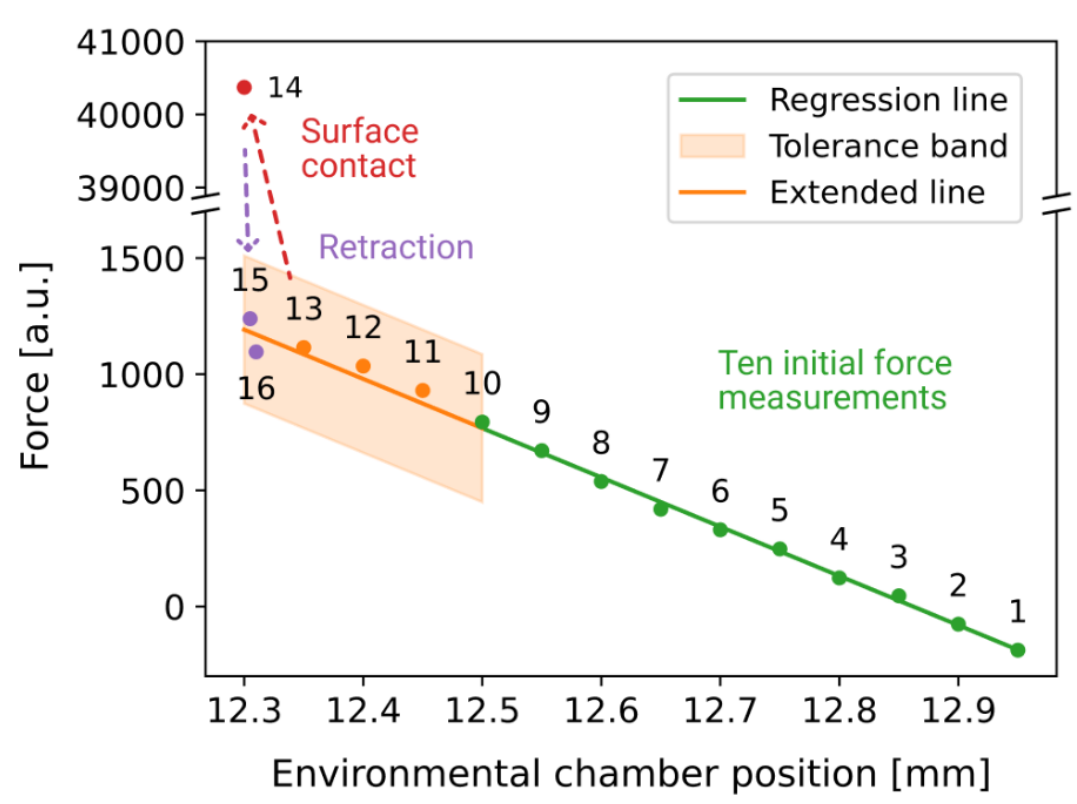

**Figure S1:** Environmental chamber design and automated surface detection. (a) Cross-sectional schematic of the 3D-printed environmental chamber. The capillary is inserted into the central tube. A humidified Ar flow is directed through a ring of radial channels that exit the top of the central tube and through a concentric ring at the bottom aperture, maintaining an inert atmosphere around the capillary. Arrows indicate the Ar flow path. (b) Representative force-position profile recorded during the automated chamber approach routine. The chamber is lowered in ten discrete steps (points 1–10) and a linear regression baseline (both visualized in green) is fitted to the recorded force values. Surface contact is identified when the incremental force increase exceeds three times the regression slope (orange points 11–14). The chamber is subsequently retracted until the force returns below the baseline (retraction indicated in purple) and the resulting position is recorded as the chamber operating height (point 16).

In order to determine the correct chamber operating height for each newly transferred materials library, an automated surface detection routine is performed after the transfer of the library onto the sample table. The chamber is lowered in ten discrete steps toward the surface of the materials library while Ar flow is active, and the force sensor reading is recorded at each position. A linear regression model is fitted to these ten force-position data points to establish a baseline response. During subsequent downward steps, the surface is considered being reached when the incremental force change between consecutive positions exceeds three times the slope of the regression line, indicating contact. The chamber is then retracted stepwise until the measured force falls back below the regression line, confirming that contact has been released. The corresponding z-position is recorded as the operating height of the environmental chamber, at which it hovers reproducibly above the surface of the materials library throughout the experiment (Figure S1b).

The thin-film materials libraries are mounted on CNC-machined polyvinyl chloride (PVC) sample holders and secured with gold-plated screws and washers (Feinmechanik, Faculty of Chemistry and Biochemistry, Ruhr University Bochum). The sample holders feature chamfered edges on the underside, ensuring self-aligning placement on the sample table even with imprecise robotic positioning. The sample holder and table design is illustrated in Figure S2. Spring-loaded, gold-plated contact pins embedded in the sample table establish electrical contact with the sample holder upon placement, grounding the materials library. A ballast plate integrated into the base of the sample holder provides the additional mass required to reliably compress the contact pins.

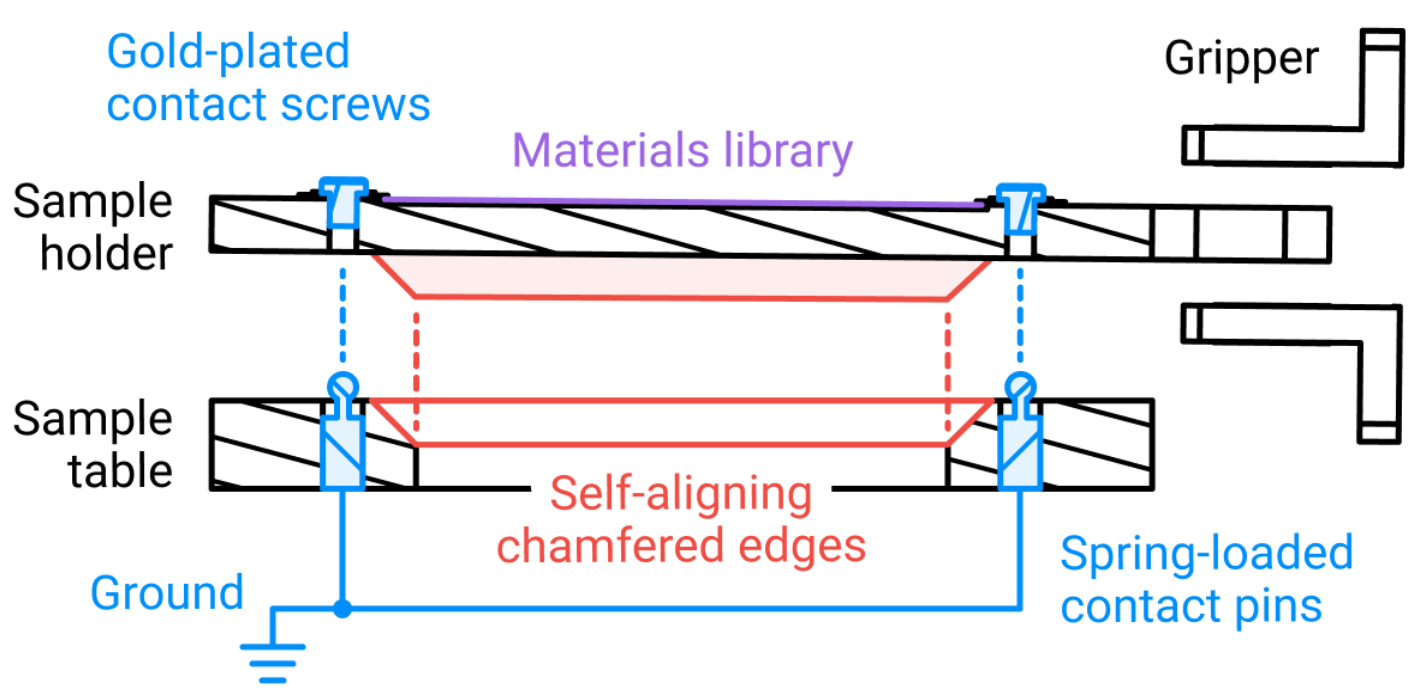

**Figure S2:** Sample holder and table design. Cross-sectional schematic of the sample holder (top) and sample table (bottom). The thin-film materials library (purple) is secured to the PVC sample holder by gold-plated screws (blue) which extend through the holder body toward its underside. Spring-loaded, gold-plated contact pins (blue) are embedded in the sample table at positions corresponding to the screws, so that placement of the holder onto the table establishes electrical contact between the materials library and the instrument ground. Chamfered edges on both components enable self-aligning placement, compensating for the positional tolerance of the robotic arm.

Sample stage translation in all three axes is achieved using a sub-micrometer-resolution stepper motor system (Owis), controlled via a four-axis LStep Express PCIe controller (Lang). A six-axis collaborative robot (UFactory Lite 6) equipped with a mechanical gripper (600 g payload) handles automated transfer of sample holders between two storage cassettes and the sample table. Each cassette accommodates five sample holders, for a total capacity of ten slots across two cassettes, all designed with the same interface geometry as the sample table. The entire instrument is enclosed within a Faraday cage to suppress electromagnetic interference.

A microscope camera (The Imaging Source, DFK 33UX264) is directed toward the capillary tip and serves a dual purpose: assisting with capillary mounting and enabling automated, real-time detection of the capillary tip and its reflection in the camera image in order to position the capillary in the working range of the piezo actuator above the sample surface. This is achieved through an automated visual approach routine shown in Figure S3. A single DC-voltage LED positioned opposite the camera, with the capillary between them, provides high-contrast backlighting. The camera image is converted to greyscale and binarized using an intensity threshold, after which a contour detection algorithm [2] implemented in OpenCV identifies the capillary tip and its mirror reflection on the sample surface. The pixel distance between the tip contour and its reflection is continuously monitored and serves as a proxy for the tip-to-sample separation (Figure S3a).

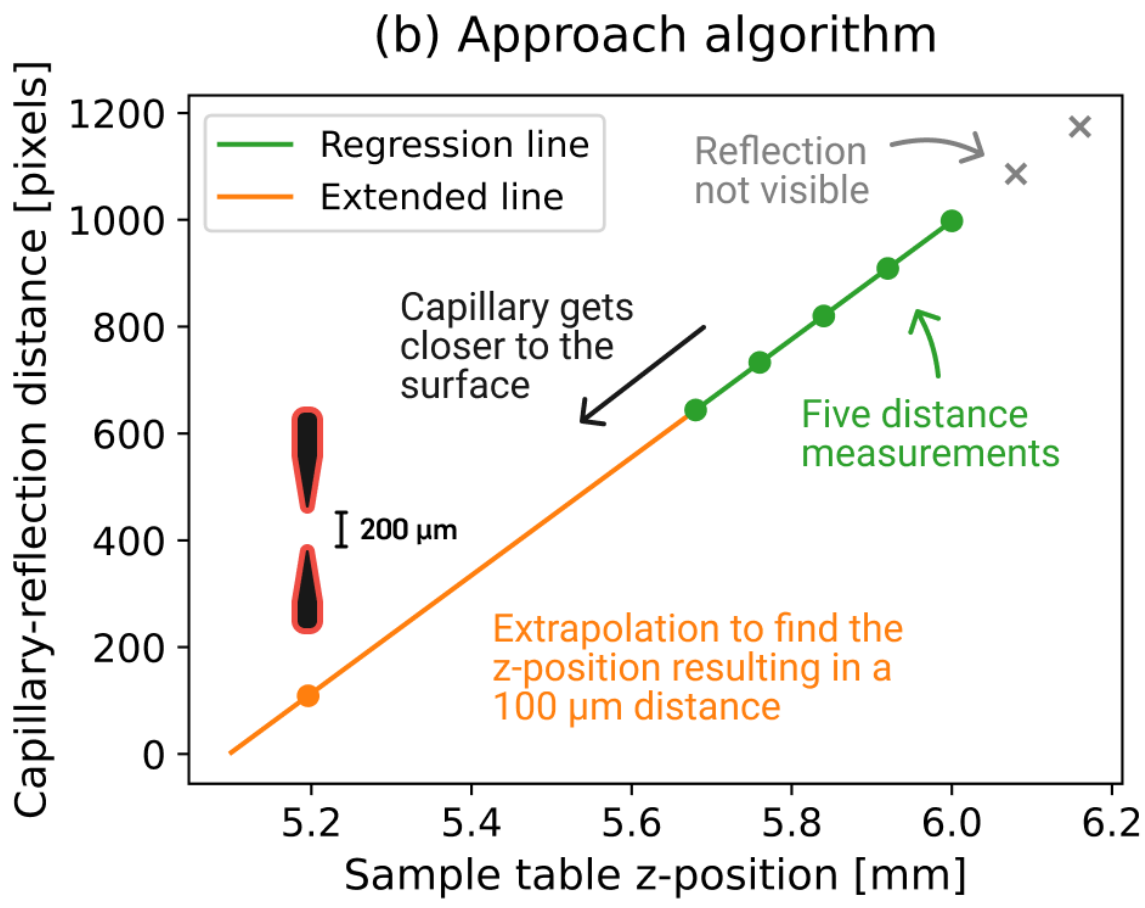

**Figure S3:** Automated visual approach routine. (a) Representative camera image showing the capillary tip and its mirror reflection on the metallic sample surface, with both contours highlighted by the detection algorithm (red outlines). (b) Illustration of the visual approach algorithm. The sample state is first raised until the capillary reflection becomes visible (grey crosses). Five discrete upward steps are then performed, and the corresponding tip-reflection pixel distances are recorded (green points). A linear regression model fitted to these data points is used to calculate the target z-position at which the tip-to-sample distance equals 100 µm, defining the handover point between the stepper motor and the piezo actuator.

The approach routine proceeds as follows (Figure S3b): The sample stage is moved upward until the reflection of the capillary tip becomes visible in the camera image. The capillary itself remains fixed in the setup, therefore only its reflection is changing during movements. The sample is afterwards moved toward the tip in five discrete steps, and the pixel distance between the tip and reflection is recorded at each position. A linear regression model is fitted to these five distance-position pairs. From the slope and intercept of the regression line, the target z-position at which the tip-to-sample distance equals 100 µm is calculated, corresponding to a tip–reflection distance in the camera image of 200 µm. The sample stage is then moved to this calculated position, placing the capillary tip within the 200 µm travel range of the piezo actuator. This strategy requires a reflective sample surface to generate a detectable mirror image of the capillary tip. All metallic thin-film materials libraries satisfy this requirement inherently. For thin films based on oxides, which are typically transparent, a metallic underlayer can be deposited to provide the necessary reflectivity. These underlayers are frequently required anyways to ensure electric conductivity of oxide films.

Due to mechanical imperfections in the instrument alignment, warp of the wafer substrates, and non-planar deformation introduced by the contact screws during mounting, a spatially varying tilt of the sample surface is likely in every experiment and must be compensated to ensure reliable capillary approach across all measurement areas. Performing the full visual approach routine described alongside Figure S3 at each measurement area is, however, not feasible. Beyond the associated time cost, the routine is susceptible to failure in areas where the capillary reflection is partially or fully obstructed, for example by dust, surface defects, or the contact screws. An assessment of the capillary visibility across all measurement areas revealed that 25 areas showed only a partial reflection of the capillary and 18 areas showed no visibility of the capillary and/or its reflection at all, predominantly at positions near the screws or the edge of the library (Figure S4a).

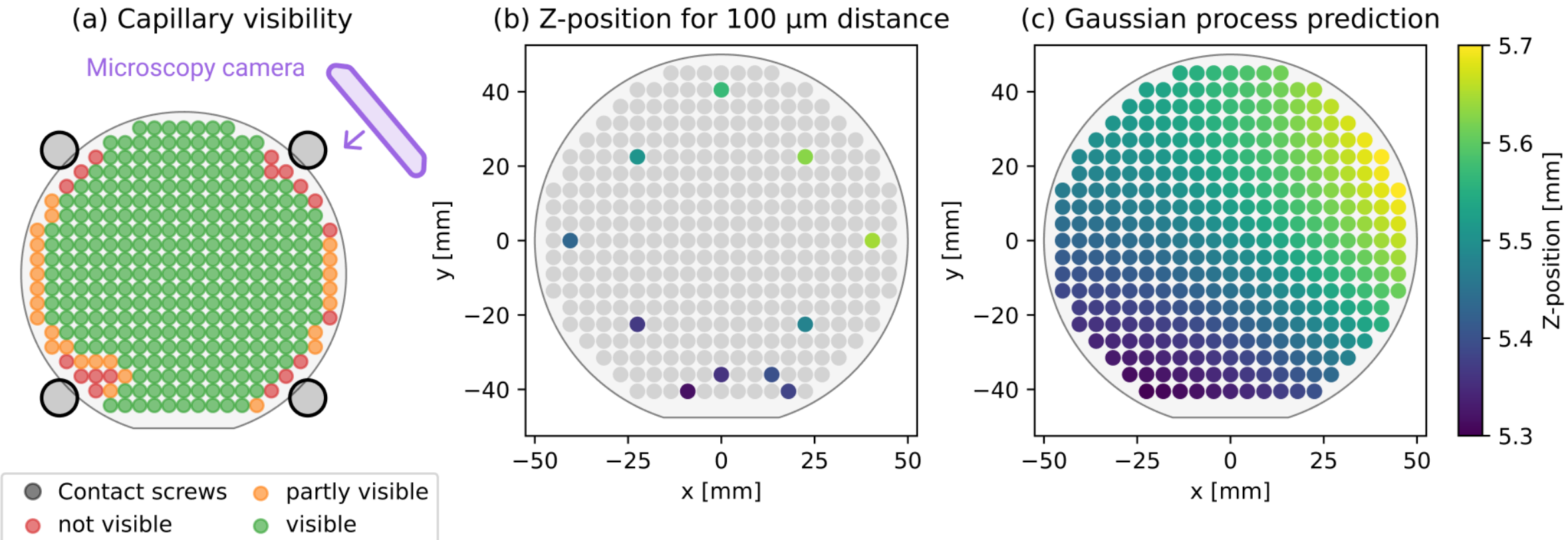

**Figure S4:** Gaussian process-based tilt correction across a materials library. (a) Map of capillary visibility in the camera image across all 342 measurement areas. Green areas indicate full visibility, orange areas indicate partial visibility, and red areas indicate positions where the tip or its reflection is obstructed, due to the edge of the wafer or obstruction by the contact screws. (b) Distribution of the 11 calibration areas selected for the visual approach routine color-coded with the measured z-position required to achieve a 100 µm tip-to-sample distance. (c) Predicted z-positions across all measurement areas obtained from a Gaussian process model trained on the 11 calibration measurements. The predicted surface profile is used to set the step motor approach position at all areas where the visual approach routine is not performed directly.

To enable a robust approach across the full library, a Gaussian process tilt correction algorithm was implemented. First, the visual approach routine is performed at 11 measurement areas distributed across the library, selected to provide broad coverage of the surface with a concentration toward the lower portion of the library, where CNC machining artifacts in the sample holder introduce greater non-planarity (Figure S4b). At each of these calibration areas, the z-position corresponding to a tip-to-surface distance of 100 µm is determined using the visual approach routine and recorded. A Gaussian process model with a squared exponential kernel is then trained on these 11 position measurements, taking the x-y coordinates of each calibration area as input and the corresponding z-positions as output training data. The trained model then predicts the required approach z-position at all remaining areas across the library (Figure S4c). A Gaussian process model was chosen over a simpler linear fit to accommodate the possibility of non-linear surface deformation and since a Gaussian process is already part of the software dependencies as part of the active learning strategy.

The control software consists of two communicating components. Fine positioning via the piezo actuator and electrochemical data acquisition via an FPGA card (PCIe-7852R, National Instruments) coupled to a variable-gain transimpedance amplifier (DLPCA-200, FEMTO Messtechnik) are handled by a LabVIEW program, extending an earlier version of the control software. Orchestration of the remaining components, including sample transfer, stepper motor control of the sample stage and environmental chamber, real-time tip detection, automatic tilt correction, and execution of the active learning algorithm, is implemented in Python and was developed specifically for this work. The two software components communicate via a REST API using a server–server architecture. The API endpoints are shown in Figure S5. Depending on the request, both interfaces are operating as a server or client. Under typical operating conditions, electrochemical characterization of a single measurement area requires approximately 90 s, while sample transfer (including environmental chamber approach and automatic tilt correction) takes approximately 4 min.

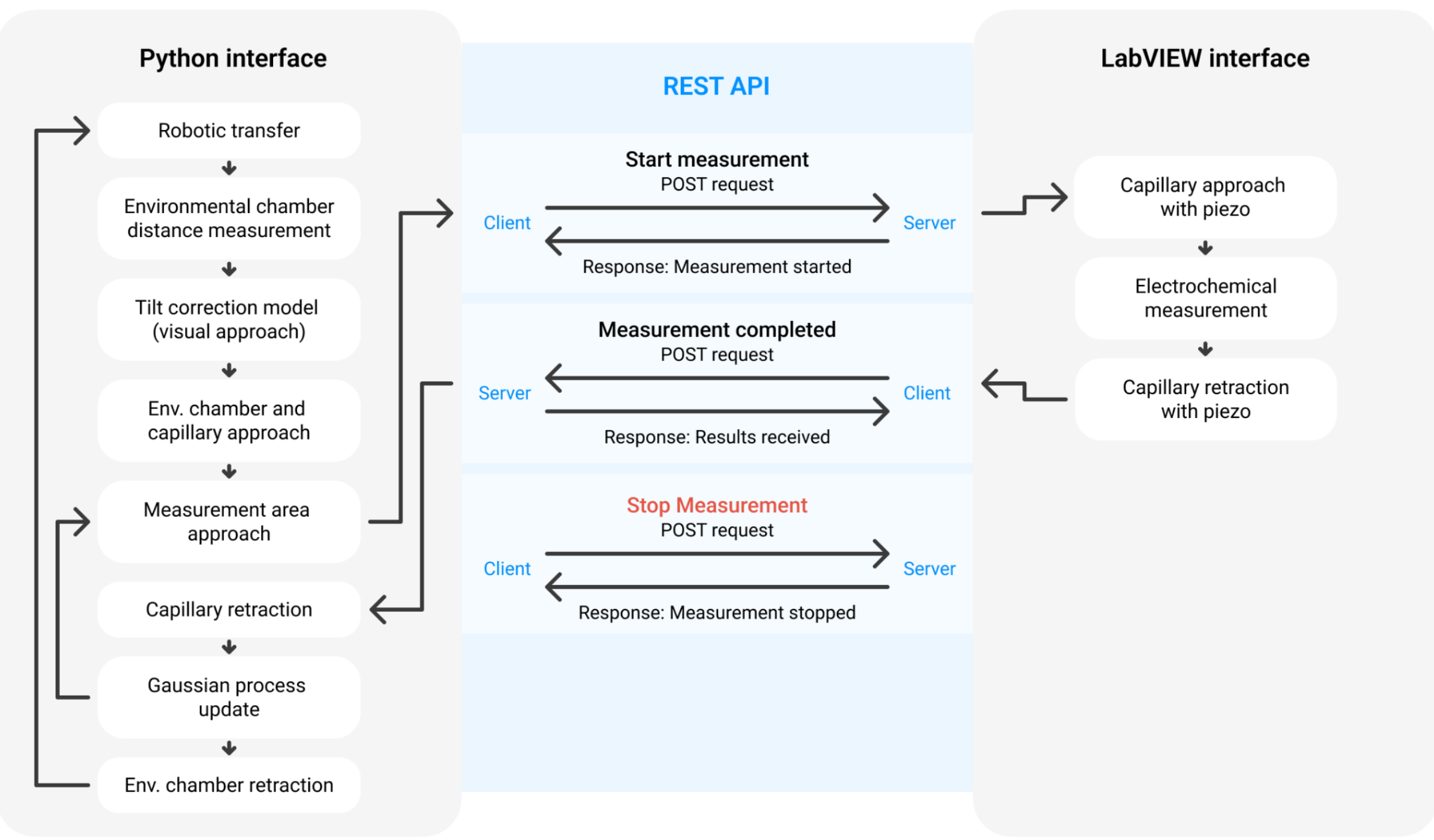

**Figure S5:** Flow diagram of the measurement procedure and the REST API between the Python and LabView interfaces. The Python interface handles the robotic transfer as well as the environmental chamber and visual capillary approach. After the capillary has reached its handoff position 100 µm, the Python interface sends a POST request to the LabView interface for starting the measurement. LabView continues with approaching the capillary using the piezo actuators and proceeds with the electrochemical measurement. Afterwards, the capillary is retracted with the piezo and a POST request indicating the completion of the measurement is sent to the Python interface, which continues by retracting the capillary with the step motors and updating the Gaussian process with the newly acquired measurement results. The environmental chamber remains at its hovering distance above the surface of the sample to ensure an inert atmosphere around the capillary. From there, the device either proceeds with measuring the next area on the current library or decides to switch the library according to the shown active learning algorithm.

### 1.7. Electrochemical measurements and data treatment

The HER voltammograms were measured in aqueous solutions containing 0.1 M $HClO_4$ and 0.1 M $LiClO_4$ (pH 1.2) with a scan rate of 1 V/s. A reversible hydrogen reference electrode (Gaskatel Mini-HydroFlex) was used. The measured voltammograms exhibited a sigmoidal shape, which is characteristic of HER voltammograms in SECCM. Although the same SECCM tip was used throughout, mass transport varies with surface wetting [3]. It depends on the material's properties and the measurement time, as evidenced by variations in the limiting current ($i_{lim}$) of the sigmoidal voltammograms.

We used the analytical expression for the steady-state SECCM voltammograms (Equation 1) [4] to extract the kinetic parameter, which is decoupled from mass transport effects.

$$i = \frac{-nFAmc_b \cdot \exp\left(-\alpha f\eta\right)}{\frac{m}{k^0} + \exp(-\alpha f\eta) + \exp\left((1-\alpha)f\eta\right)} \quad (1)$$

Here $n$ is the number of the transferred electrons ($n = 2$), $F$ is the Faraday constant, $m$ is the mass transport coefficient (defined as $m = i_{lim}^{SECCM}/nFAc_b$), $\alpha$ is the charge transfer coefficient, $\eta$ is the overpotential (defined as $\eta = E - E^0$), and $k^0$ is the standard rate constant. Each LSV was fitted with the fitting variables of $i_{lim}$, $\alpha$ and $k^0$ with non-linear least-squares regression. The individual effects of the three parameters on the LSV's shape are illustrated in Figure S6. $k^0$ represents intrinsic electrocatalytic kinetics, while $\alpha$ can be interpreted as a mechanistic indicator in HER, given that the Tafel slope $= 2.303\ RT/\alpha F$ in the Tafel region. However, we note that $\alpha$ values extracted from SECCM voltammograms should be regarded as apparent parameters rather than accurate mechanistic interpreters. The high scan rates in SECCM voltammetry may distort the $\alpha$ values.

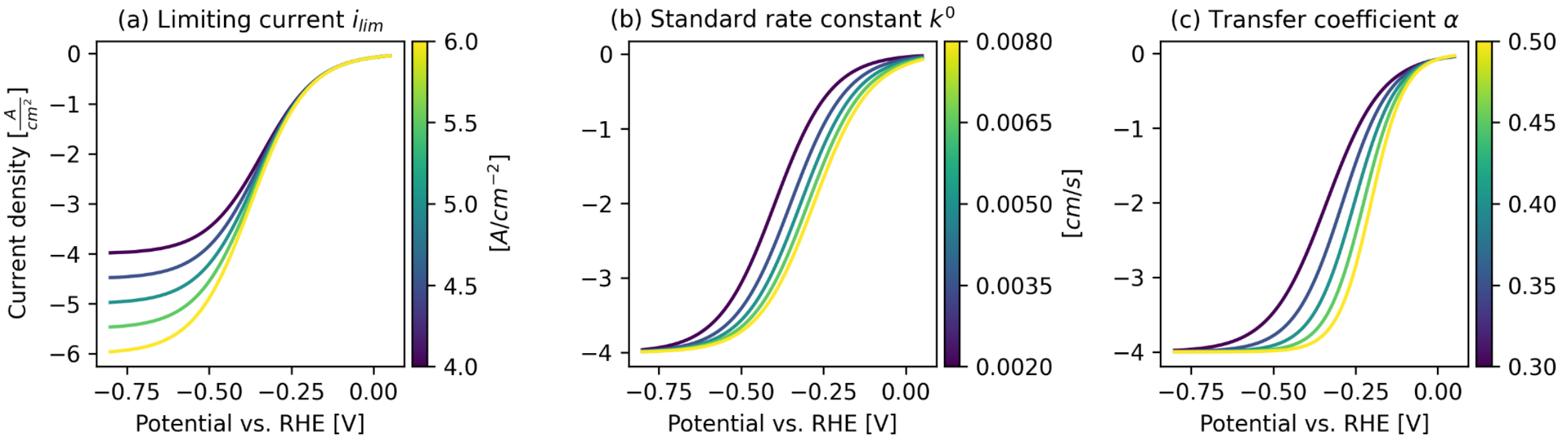

**Figure S6:** Effect of the LSV parameters on the LSV's shape with the standard parameters $i_{lim} = 4$, $k^0 = 0.004$, and $\alpha = 0.3$. In each subplot, one parameter is varied according to the values indicated on the colorbars, while the other parameters are kept constant. (a) Change of the shape when varying the limiting current $i_{lim}$, which controls the minimum current density value the LSV asymptotically approaches with higher absolute potentials. (b) Effect of the standard rate constant $k^0$ on the shape of the LSV. $k^0$ controls the position of the inflection point on the potential axis (c) Effect of the transfer coefficient $\alpha$ on the LSV. $\alpha$ controls the symmetry of the LSV.

### 1.8. Cost-aware active learning framework

To predict the transfer coefficient $\alpha$ and standard rate constant $k^0$ from the volume compositions, a multi-output Gaussian process with coregionalization kernel implemented in GPflow [5] was used. A Gaussian process models a function $f(x)$ as a distribution over functions, such that any finite set of function values follows a multivariate normal distribution:

$$f(x) \sim GP(\mu(x), k(x, x'))$$

where $\mu(x)$ is the mean function and $k(x, x')$ is the covariance function [6]. The mean function is commonly set to $\mu(x) = 0$, as the input training data can be standardized, and the Gaussian process is sufficiently flexible to capture the underlying mean structure. The covariance function is defined by a kernel function which encodes the correlation between function values at inputs $x$ and $x'$ and therefore controls the function's shape. Popular kernel choices are the squared exponential kernel and the Matérn kernel class [7]. Due to preliminary studies [8,9] of Gaussian processes predicting electrochemical activity and other functional properties of thin-film materials libraries based on compositional trends, a Matérn kernel with $\nu = 5/2$ was used.

Analysis of $\alpha$ and $k^0$ distributions across materials libraries measured with long-range SECCM before the development of the autonomous SECCM platform revealed that $k^0$ consistently varies over several orders of magnitude within a single library. A logarithmic transformation of $k^0$ was therefore applied prior to the Gaussian process modeling to improve the predictive accuracy.

Since $\alpha$ and $k^0$ are jointly extracted from a single LSV fit, they are inherently correlated. Modeling each output with an independent Gaussian process would fail to capture this structure. A coregionalization kernel [10] was therefore used, which augments the input-space kernel with an output covariance matrix $K_f$ encoding the inter-output relationships:

$$K_{coregion}((x,l),(x',l')) = K_f(l,l') \cdot k(x,x')$$

where $l$ and $l'$ denote the different outputs. The Gaussian process is trained on the volume compositions as inputs and the (partially transformed) parameters extracted from the LSVs of the already measured areas. The compositions are converted to isometric log-ratio (ILR) transformed, which resolves the sum-to-one constraint of the raw compositions [11,12]. At each area, the Gaussian process posterior is tasked to predict the mean and epistemic uncertainty. Predicted $k^0$ values are back transformed via exponentiation for interpretation. The uncertainty is used as the acquisition function for pure-exploration active learning. To determine when to switch libraries, the epistemic uncertainty is weighted by the estimated measurement time $t_{meas}$ for each measurement area, computed with:

$$t_{meas} = t_{trans} + t_{acq} + d_{area}/v_{step}$$

Where $t_{trans}$ is the fixed sample transfer time, $t_{acq}$ is the data acquisition time, $d_{area}$ is the distance from the current position to the candidate area, and $v_{step}$ is the lateral velocity of the stepper motors. The area with the highest acquisition-cost-weighted epistemic uncertainty is selected as the next measurement target.

## 2. Supplementary figures (Figure S7–S23) and supplementary notes

### 2.1. Analysis of the volume composition

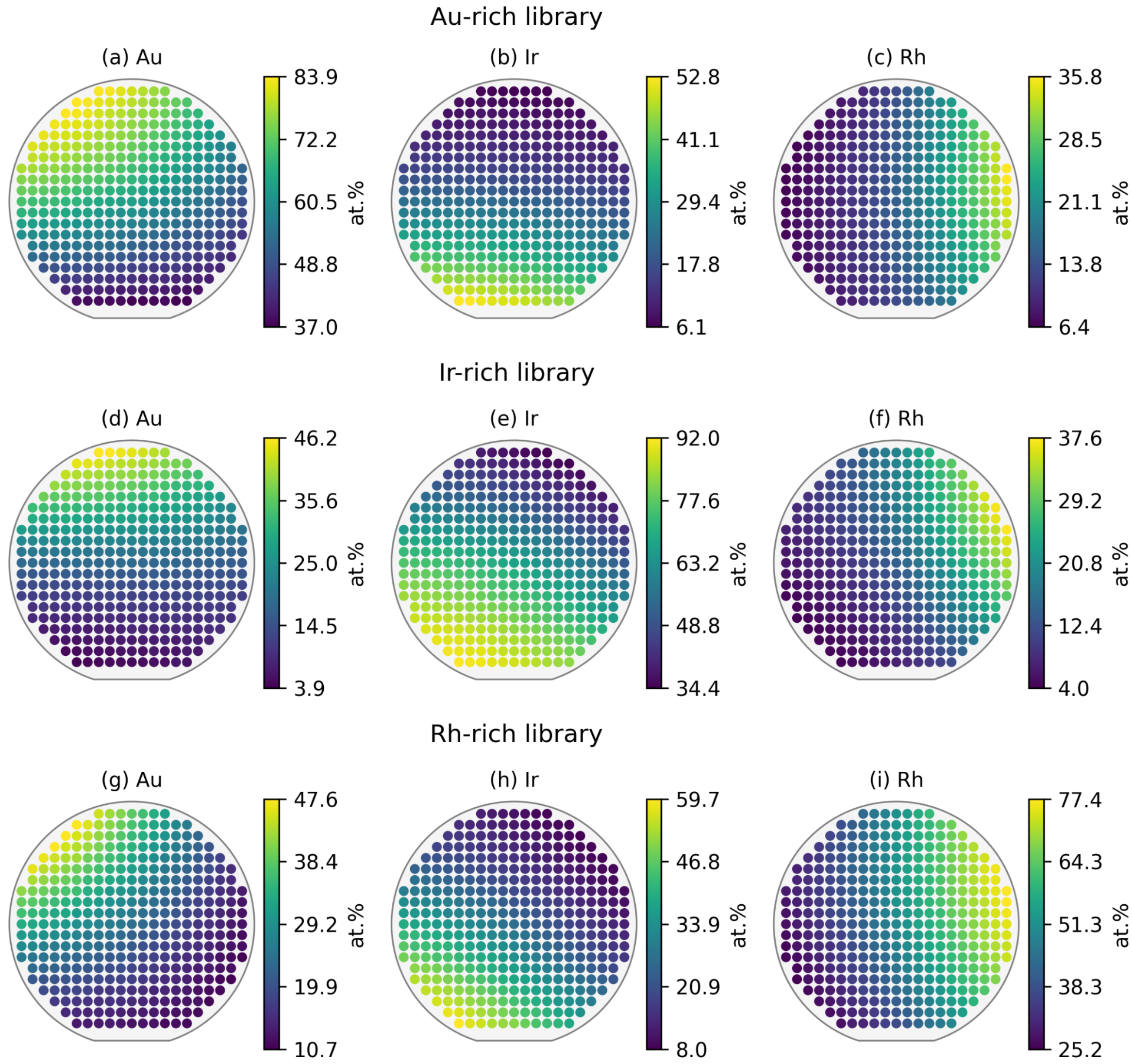

**Figure S7:** (Volume) compositions of the Au-rich (a-c), Ir-rich (d-f) and Rh-rich (g-i) libraries acquired by EDX.

### 2.2. Analysis of the surface composition

The autonomous exploration of the Au–Ir–Rh system was performed using compositions acquired by energy dispersive X-ray spectroscopy (EDX). EDX is a high-throughput technique which enables automated compositional mappings of full thin-film materials libraries in approximately ~1.5 h, making it a practical and cost-effective method for large-scale screening. However, as the probing depth of EDX extends into the micrometer range, the measured compositions represent volume-averaged compositions and can differ from the true surface compositions, e.g. which may arise due to surface segregation effects [13]. Since electrochemical activity is governed by the topmost one to three atomic layers of the catalyst [14], it is important to assess whether EDX-derived compositions can serve as a reliable proxy for the surface composition. Therefore, complementary surface-sensitive (approximately 5-10 nm information depth) X-ray photoelectron spectroscopy (XPS) measurements were performed at 13 evenly distributed areas on each of the three libraries. Given a measurement duration of ~1.5 h per composition, full library mappings by XPS are generally experimentally not feasible.

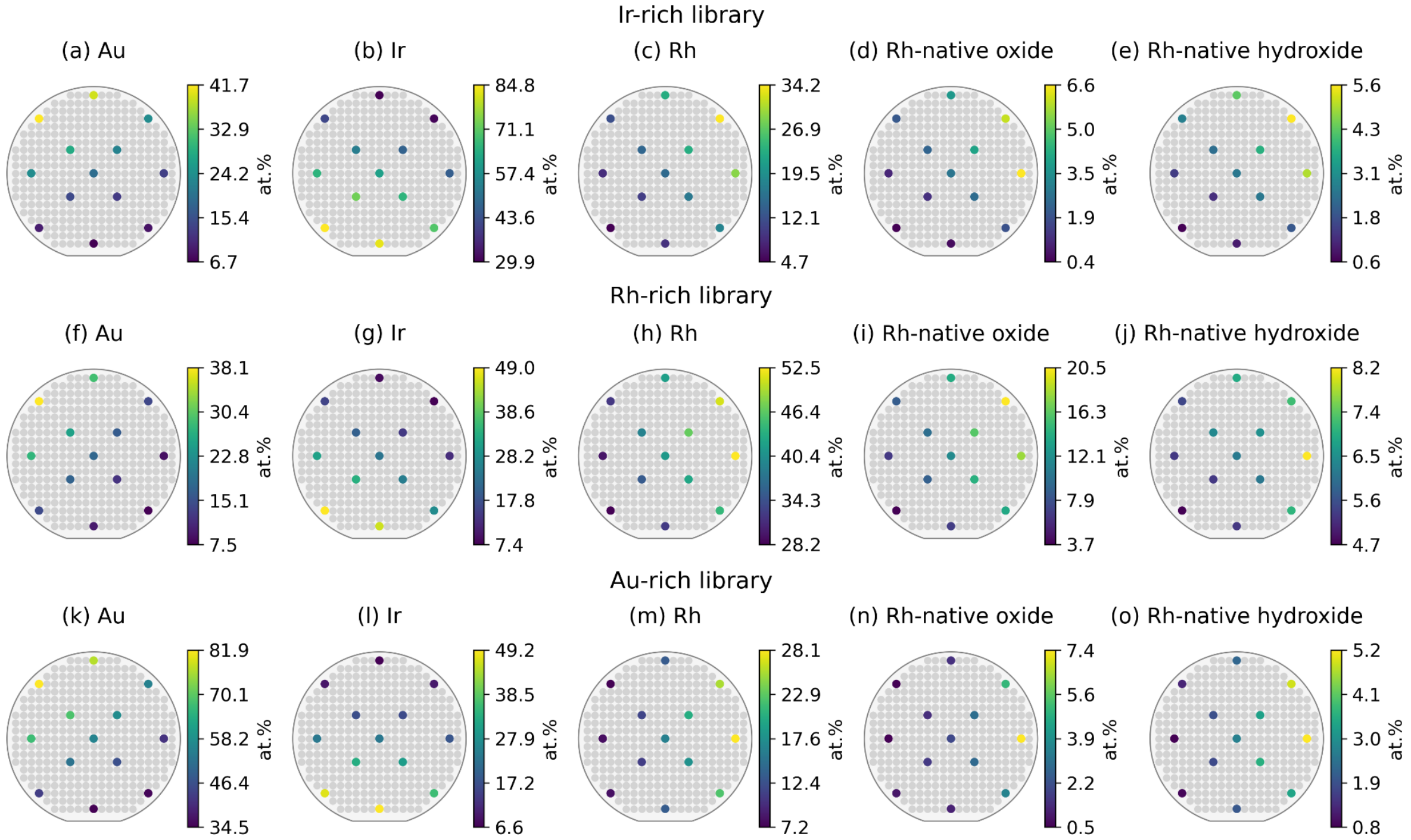

**Figure S8:** Surface compositions determined by XPS at 13 evenly distributed areas per library, displayed on the 342-measurement area grid. The results are arranged in three rows corresponding to the three libraries, with columns showing the contents of metallic Au, Ir, Rh, native Rh-oxide and Rh-hydroxide in at.%. The spatial gradients of the metallic Rh, Rh-oxide and Rh-hydroxide contents follow similar trends across all libraries.

The XPS results are shown in Figure S8, displayed on the same 342-measurement area grid used throughout this work. In addition to the metallic peaks of Au, Ir, and Rh, detectable native Rh oxide and hydroxide peaks were identified in the XPS spectra. The gradient of both the Rh-oxide and Rh-hydroxide content follows that of the metallic Rh, consistent with a composition-dependent oxidation behavior.

Figure S9 shows the correlation between surface and volume compositions for all three constituents, with the black dashed line indicating perfect correlation. For this analysis, the metallic and oxide Rh contributions were summed to obtain the total surface Rh content. The correlation points for all three libraries fall close to the perfect correlation line, demonstrating that the surface compositions are in good agreement with the volume compositions across the full compositional range. The mean Euclidean distance between surface and volume compositions is 4.2 at.%. This is consistent with the magnetron co-sputtering fabrication process at room temperature, as under these conditions, sputtered atoms are quenched upon arrival at the substrate, suppressing diffusion. A minor surface depletion of Ir (and a corresponding enrichment of Rh) is observed at contents of 40-50 at.%. The fact that the correlation trends extend continuously across library boundaries shows the reproducibility of library synthesis by magnetron co-sputtering.

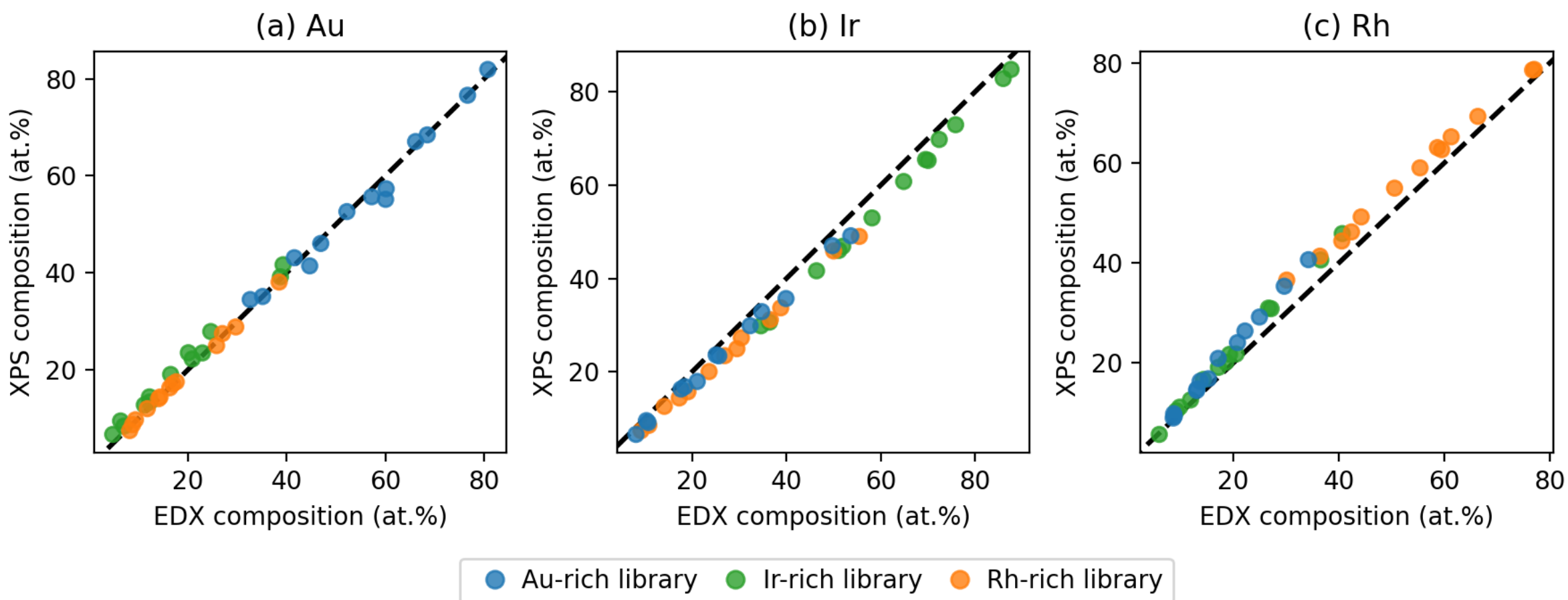

**Figure S9:** Correlation between XPS-derived surface compositions and EDX-derived volume compositions for (a) Au, (b) Ir, and (c) Rh across the libraries. Data points are color-coded by library. The black dashed line indicates perfect agreement between surface and volume compositions. For Rh, the metallic and native oxide contents were summed to obtain the total surface Rh content. Data points from all libraries fall close to the perfect correlation line across the full compositional range. A minor surface depletion of Ir and corresponding enrichment of Rh is observed.

To obtain surface composition estimated for the full set of measurement areas beyond the 39 XPS-characterized locations, the remaining surface compositions were predicted from the volume compositions using multi-output Gaussian process regression following the approach described in [8]. Figure S10a shows the resulting overlay of volume and predicted surface compositions across the full experimentally covered compositional range. The distribution is consistent with the pairwise correlation shown in Figure S9, showing a minor shift of the surface compositions toward Rh-rich regions relative to the volume compositions. Figure S10b shows the spatial distribution of the surface Rh oxide content across the composition space. The oxide fraction increases with Rh content, reaching nearly 30 at.% in the most Rh-rich areas, and is negligible at Rh contents below approximately 10-20 at.%. Additionally, the Rh oxide fraction is systematically higher in Au-Rh-rich regions than in Ir-Rh-rich ones, suggesting a preferential oxidation of Rh when combined with Au.

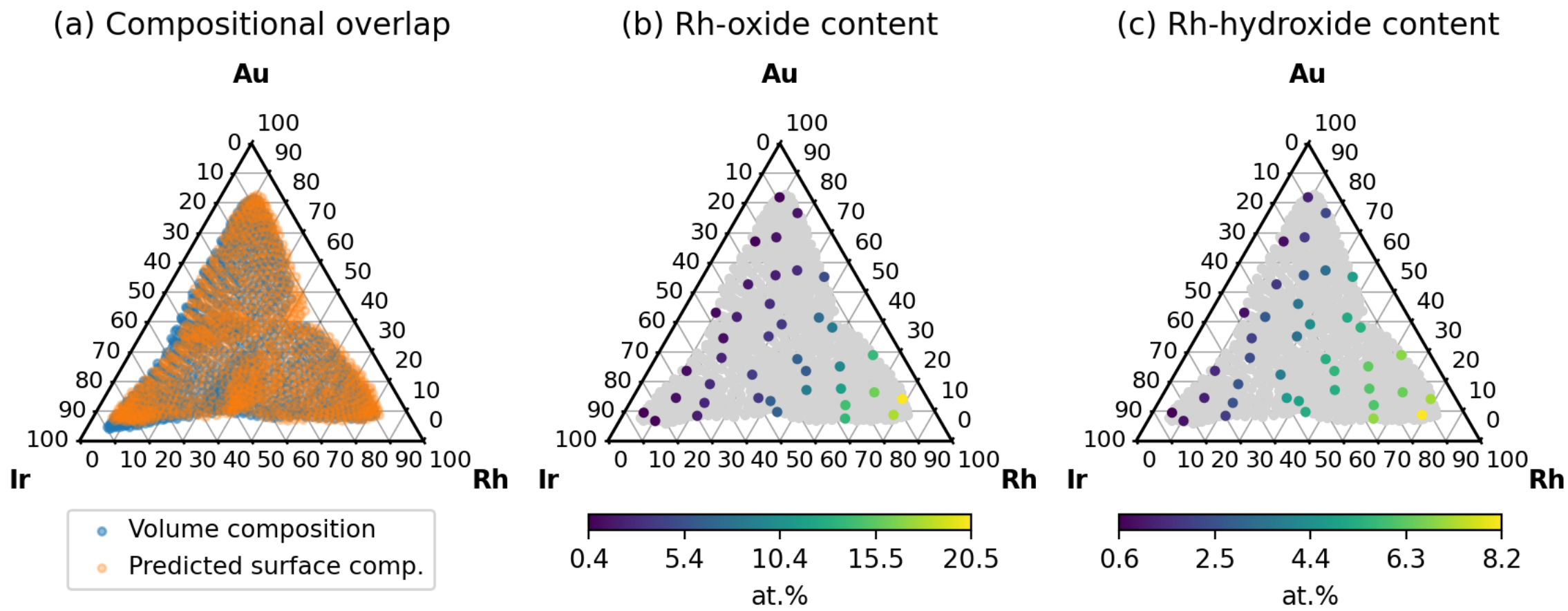

**Figure S10:** Comparison of measured volume composition (EDX) and Gaussian process regression predicted surface compositions across the full Au–Ir–Rh composition space. (a) Overlay of predicted surface compositions and volume compositions on the ternary composition space for all measurement

areas. The systematic shift of surface compositions toward Rh-rich regions is consistent with Figure S9. (b) Native Rh-oxide content mapped onto the ternary composition space. The Rh-oxide fraction increases with total Rh content, reaching a maximum of 20 at.%, and is negligible at Rh contents below approximately 50 at.%. (c) Native Rh-hydroxide content mapped onto the composition space. Consistent with the Rh-oxide content, the hydroxide increases with Rh-content as well but starts to form at Rh-contents of ~30 at.% or higher. The highest measured Rh-hydroxide content is 8.2 at.%.

## 2.3. Analysis of the phase constitution

XRD was performed on all three Au–Ir–Rh thin-film materials libraries to verify that a single-phase solid solution is present across the full compositional range covered by the libraries. This validation is necessary to ensure that the variations in electrocatalytic HER activity observed by the SECCM can be attributed to compositional effects rather than to differences in phase constitution or crystal structure.

Figures S11, S12, and S13 summarize the XRD results for the Au-rich, Ir-rich, and Rh-rich libraries, respectively. Each figure is structured identically: subplot (a) of each figure shows three representative diffractograms recorded on measurement areas 1 (Ir-rich), 241 (Rh-rich) and 312 (Au-rich) compositions (positions indicated in the inset), together with the reference peak positions of the dominant element as vertical lines; subplot (b) shows the distribution of the fitted 2θ positions for each of the five reflections across the full library. Peak positions were extracted by fitting a Voigt function to each reflection with non-linear least-squares regression.

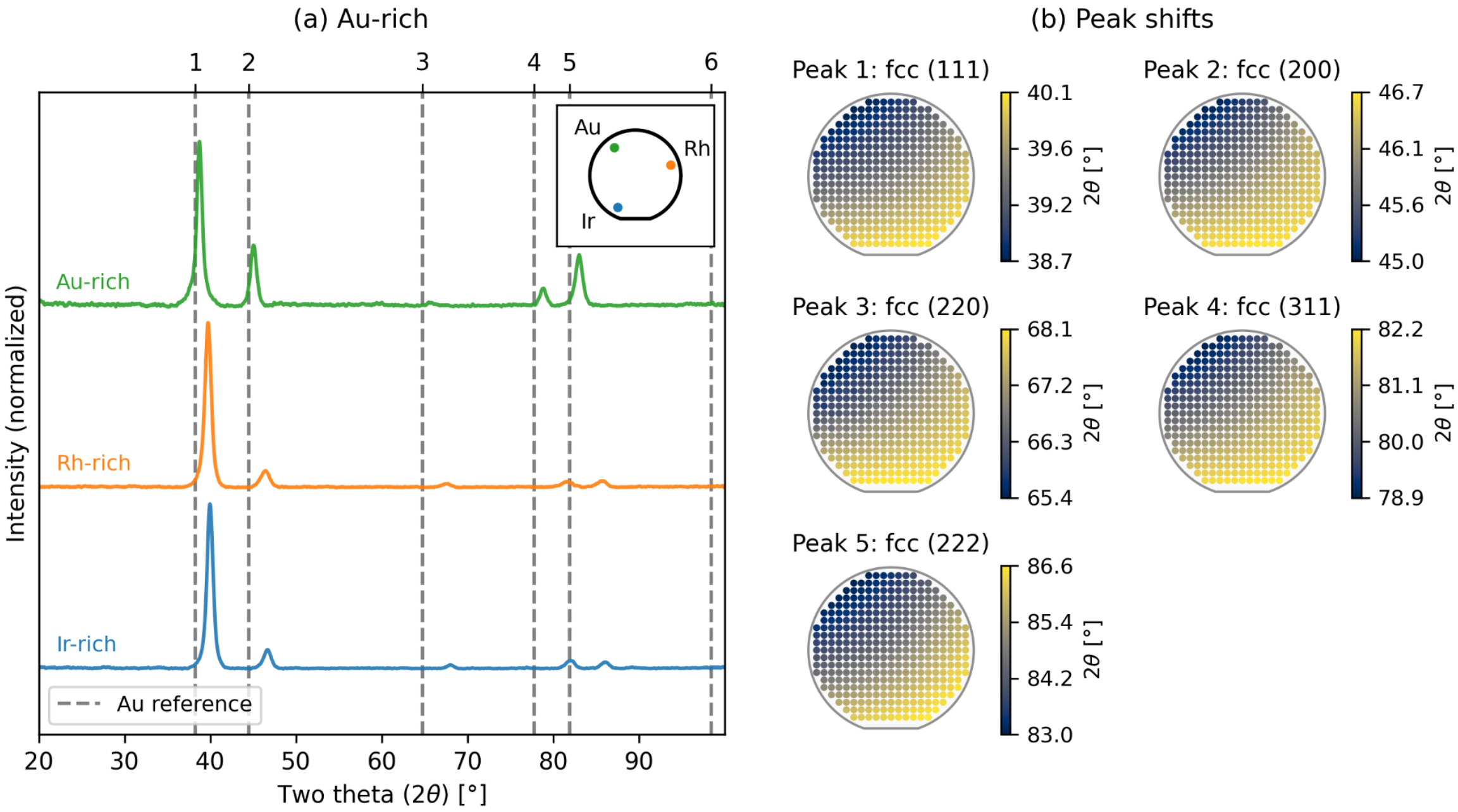

**Figure S11:** XRD analysis of the Au-rich materials library. (a) Representative diffractograms recorded at three measurement areas with Au-rich (measurement area 312), Ir-rich (area 1), and Rh-rich (area 241) compositions within the library. The inset indicates the location of the selected measurement areas on the wafer level of the library, which were selected based on the cathode positions in the deposition chamber. The vertical dotted lines mark the reference peak positions of fcc Au (ICSD 53764). (b) Distributions of the fitted 2θ peak positions across the library for each of the five reflections assigned to the (111), (200), (220), (311), and (222) planes of the fcc lattice. Peak positions were extracted by non-linear least-squares fitting of a Voigt function to each reflection.

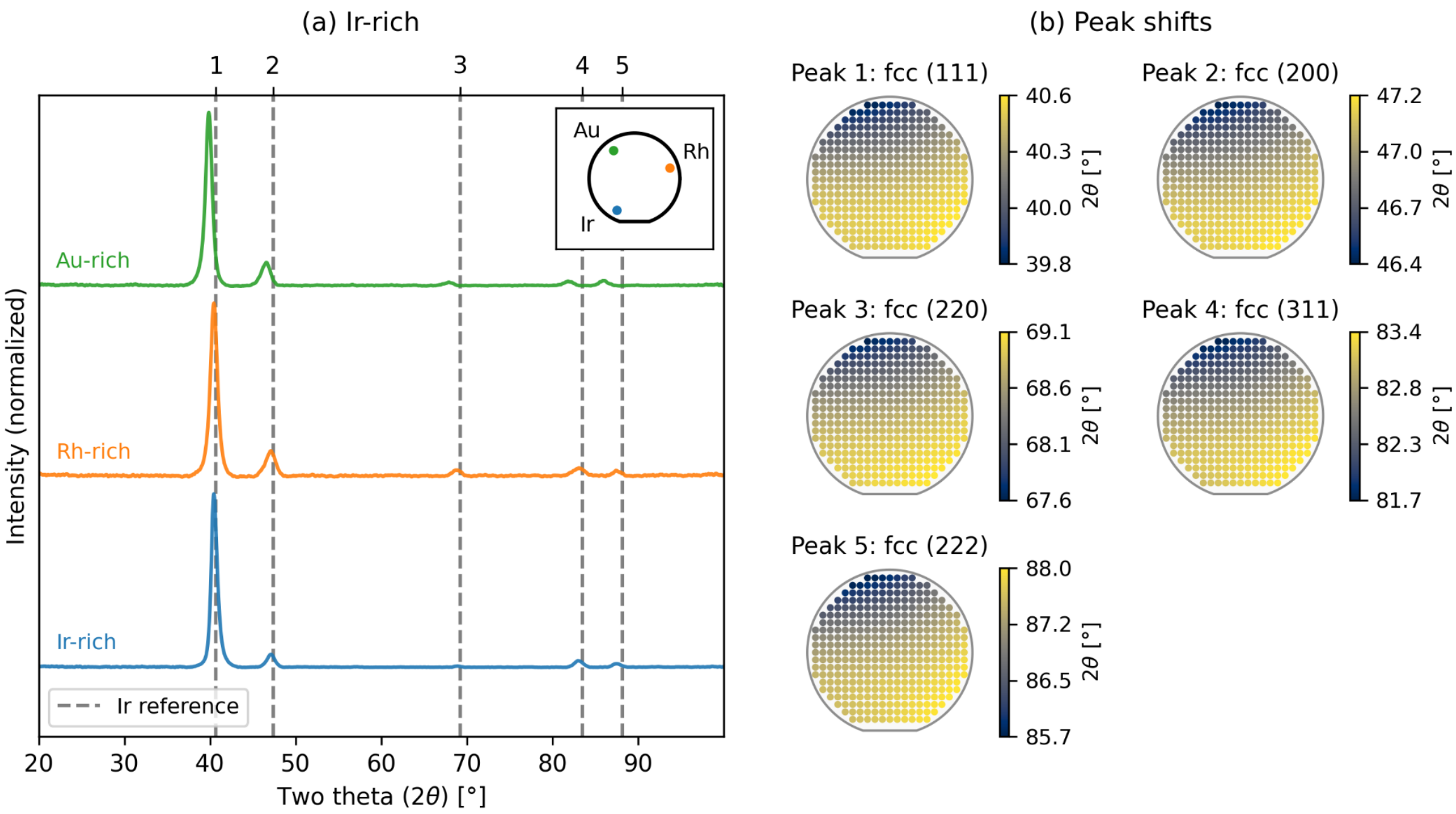

**Figure S12:** XRD analysis of the Ir-rich materials library. (a) Representative diffractograms recorded at three measurement areas with Au-rich (measurement area 312), Ir-rich (area 1), and Rh-rich (area 241) compositions within the library. The inset indicates the location of the selected measurement areas on the physical library, which were selected based on the cathode positions in the deposition chamber. The vertical dotted lines mark the reference peak positions of fcc Ir (ICSD 64992). (b) Distributions of the fitted 2θ peak positions across the library for each of the five reflections assigned to the (111), (200), (220), (311), and (222) planes of the fcc lattice. Peak positions were extracted by non-linear least-squares fitting of a Voigt function to each reflection.

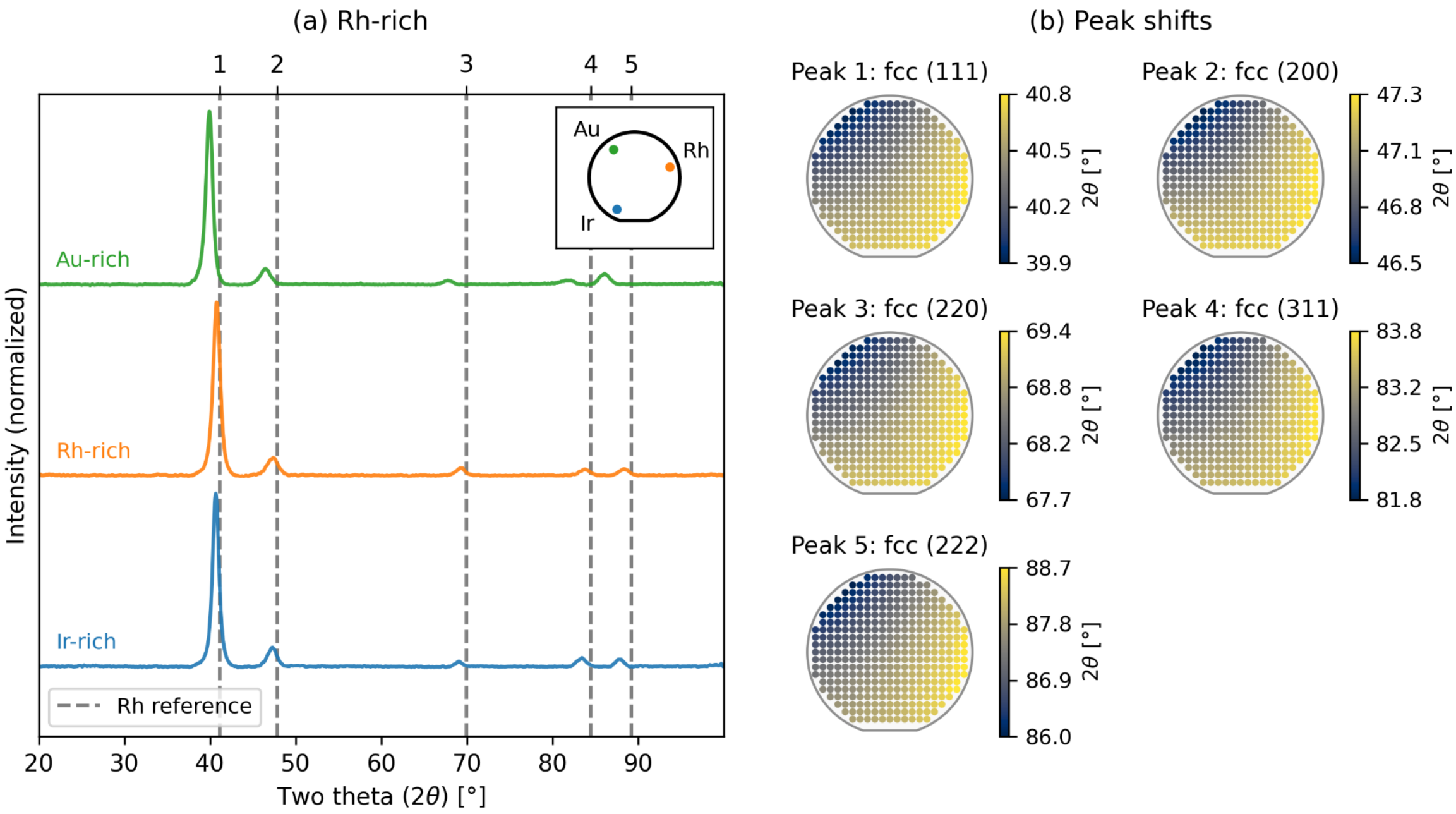

**Figure S13:** XRD analysis of the Rh-rich materials library. (a) Representative diffractograms recorded at three measurement areas with Au-rich (measurement area 312), Ir-rich (area 1), and Rh-rich (area 241) compositions within the library. The inset indicates the location of the selected measurement areas on the physical library, which were selected based on the cathode positions in the deposition chamber.

The vertical dotted lines mark the reference peak positions of fcc Rh (ICSD 52064). (b) Distributions of the fitted 2θ peak positions across the library for each of the five reflections assigned to the (111), (200), (220), (311), and (222) planes of the fcc lattice. Peak positions were extracted by non-linear least-squares fitting of a Voigt function to each reflection.

In all three libraries, five reflections are consistently observed around 2θ of 40°, 46.7°, 68°, 82.2° and 86.5°, respectively, which correspond to the (111), (200), (220), (311) and (222) planes of an fcc lattice. No additional reflections are detected, confirming that a single-phase solid solution across the full measured compositional range is present. All five peaks across the three libraries follow the same smooth peak shift distribution from the upper left (Au-rich) to the lower right (Ir-Rh-rich). This smooth distribution also extends across libraries, as indicated in Figure S14 showing the fitted 2θ position of the (111) reflection plotted on the Au–Ir–Rh composition space. With the Au-rich corner ($c_{Au} = 83.8\ at.\%$) at 38.7° 2θ, the Ir-rich corner ($c_{Ir} = 92\ at.\%$) at 40.5° 2θ, and the Rh-rich corner ($c_{Rh} = 77.4\ at.\%$) at 40.8° 2θ, a substantially larger shift along the Au direction compared to a small shift between the Ir-rich and Rh-rich corners can be observed. This reflects the difference in the elemental lattice parameters (Au: 0.4073 nm, Ir: 0.3839 nm, Rh: 0.3803 nm [1]) and is consistent with Vegard's law for the continuous solid solution.

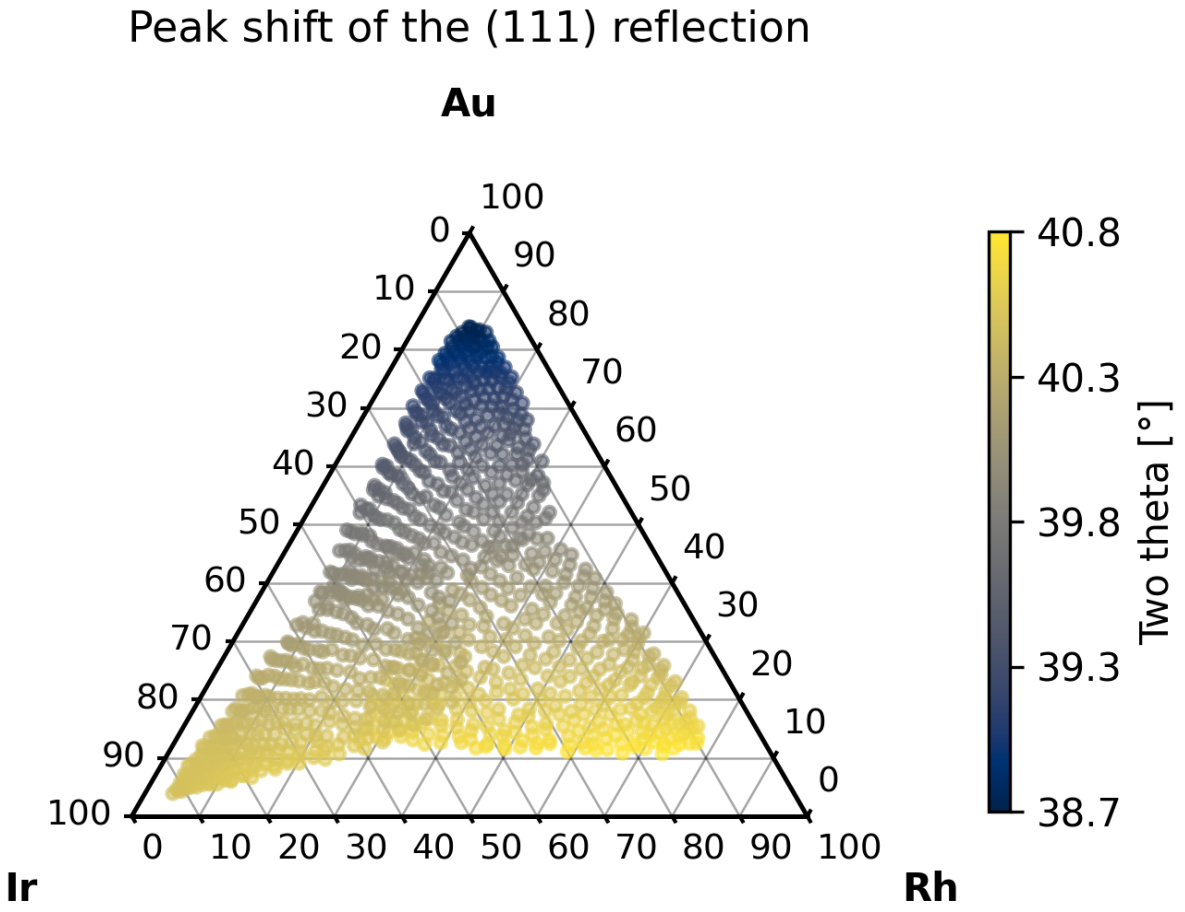

**Figure S14:** Fitted 2θ position of the (111) reflection plotted across the Au–Ir–Rh composition space, combining data from all three libraries. The peak position varies smoothly and continuously across the fabricated compositional range, with a pronounced shift along the Au direction and only a minor shift between the Ir-rich and Rh-rich corners, consistent with Vegard's law. The continuous gradient confirms the formation of a single-phase solid solution across the full investigated composition space.

In addition to the 1D diffractograms, the 2D patterns recorded from the area detector can be used to assess crystallographic texture across the composition space. Representative 2D patterns are shown in Figure S15. The Au-rich composition (Figure S15a) exhibits pronounced spots in addition to the Debye-Scherrer rings, indicating that the single-phase solid solution is textured in this region of the composition space. A similar but less pronounced effect can be observed for the Ir-rich compositions (Figure S15b), where the rings dominate and only weak spots are visible. In contrast, the Rh-rich compositions (Figure S15c) and compositions in the center of the composition space (Figure S15d) show uniform Debye-Scherrer rings only, indicating the absence of preferred orientation. Texture in the Au–Ir–Rh libraries is therefore confined to the corners of the composition space, while the films in the center regions of the composition space grow without preferred orientation.

[1] ICSD 53764, ICSD 64992, ICSD 52064

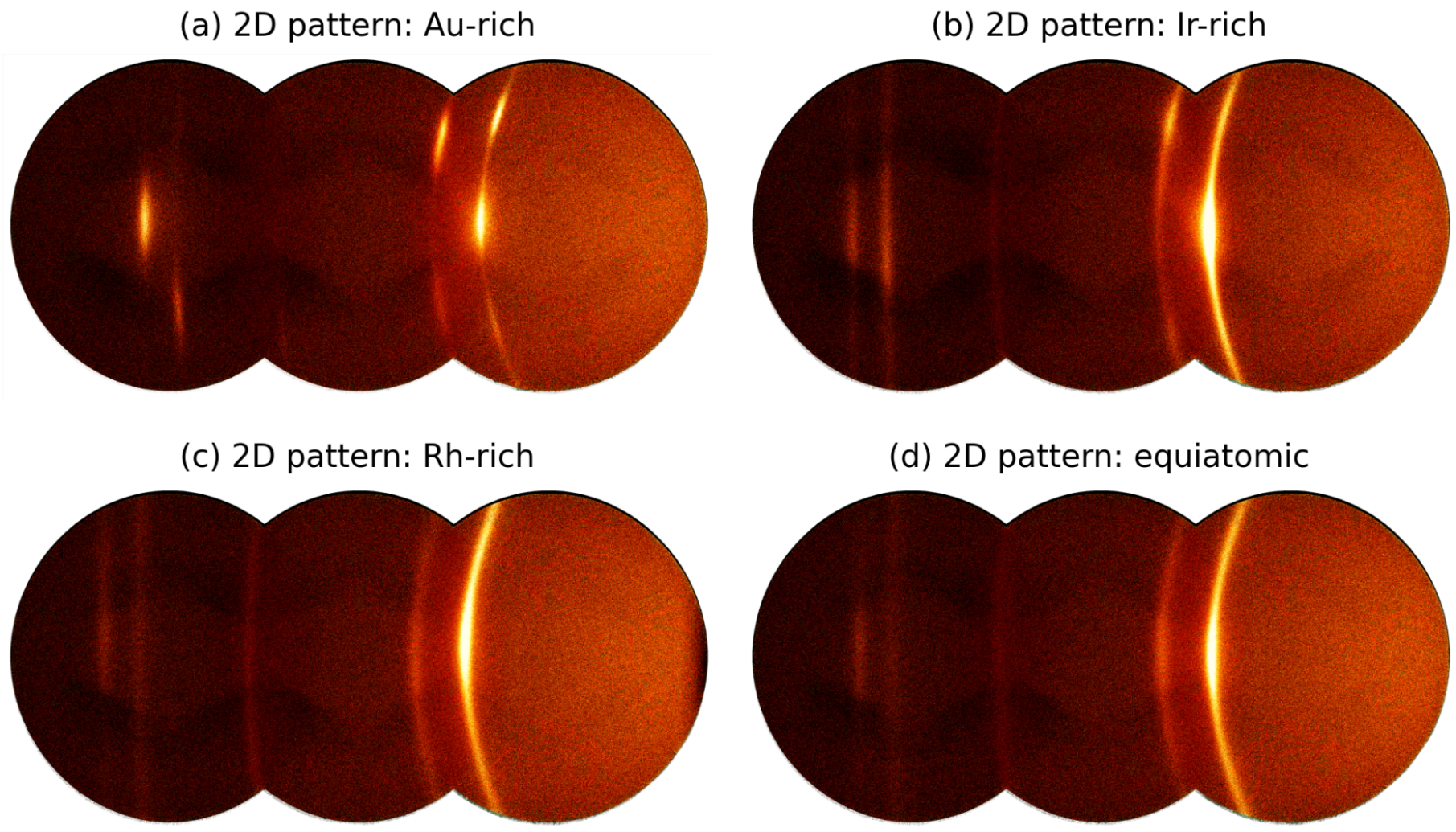

**Figure S15:** 2D XRD patterns recorded from the area detector for representative compositions across the Au–Ir–Rh composition space. (a) Au-rich composition (area 312 on the Au-rich library), showing pronounced spots together with the Debye–Scherrer rings, indicative of crystallographic texture. (b) Ir-rich composition (area 1 on the Ir-rich library), showing weak spots on otherwise dominant rings, (c) Rh-rich composition (area 241 on the Rh-rich library) and (d) composition close to the equiatomic center of the composition space (area 200 on the Rh-rich library), both showing only Debye-Scherrer rings, indicating the absence of preferred orientation.

## 2.4. Analysis of the surface morphology

In order to assess whether the observed electrocatalytic activity trends across the Au–Ir–Rh composition space could be confounded by systematic variations of the electrochemical surface area, AFM measurements were performed in 15 selected areas distributed across the three materials libraries. The AFM-derived surface roughness serves as a proxy for relative differences in surface area between compositions. A 500 nm x 500 nm area was imaged at each position, and the arithmetic mean roughness $R_a$was extracted.

The 15 measurement areas were selected to systematically sample distinct regions of the composition space (Figure S16a). Regions 1-3 were positioned at the compositionally extreme regions of the libraries, capturing the highest content of Au, Ir, and Rh, respectively. Regions 4-6 were placed at the center of each individual library. Regions 7-9 were positioned in compositional overlap regions between pairs of libraries, where two libraries have similar elemental composition but were fabricated under different deposition conditions. These regions enable a direct comparison of surface morphology across libraries. Finally, region 10 corresponds to the near-equiatomic center of the Au–Ir–Rh composition space. In this region, one AFM measurement was performed on each of the three libraries.

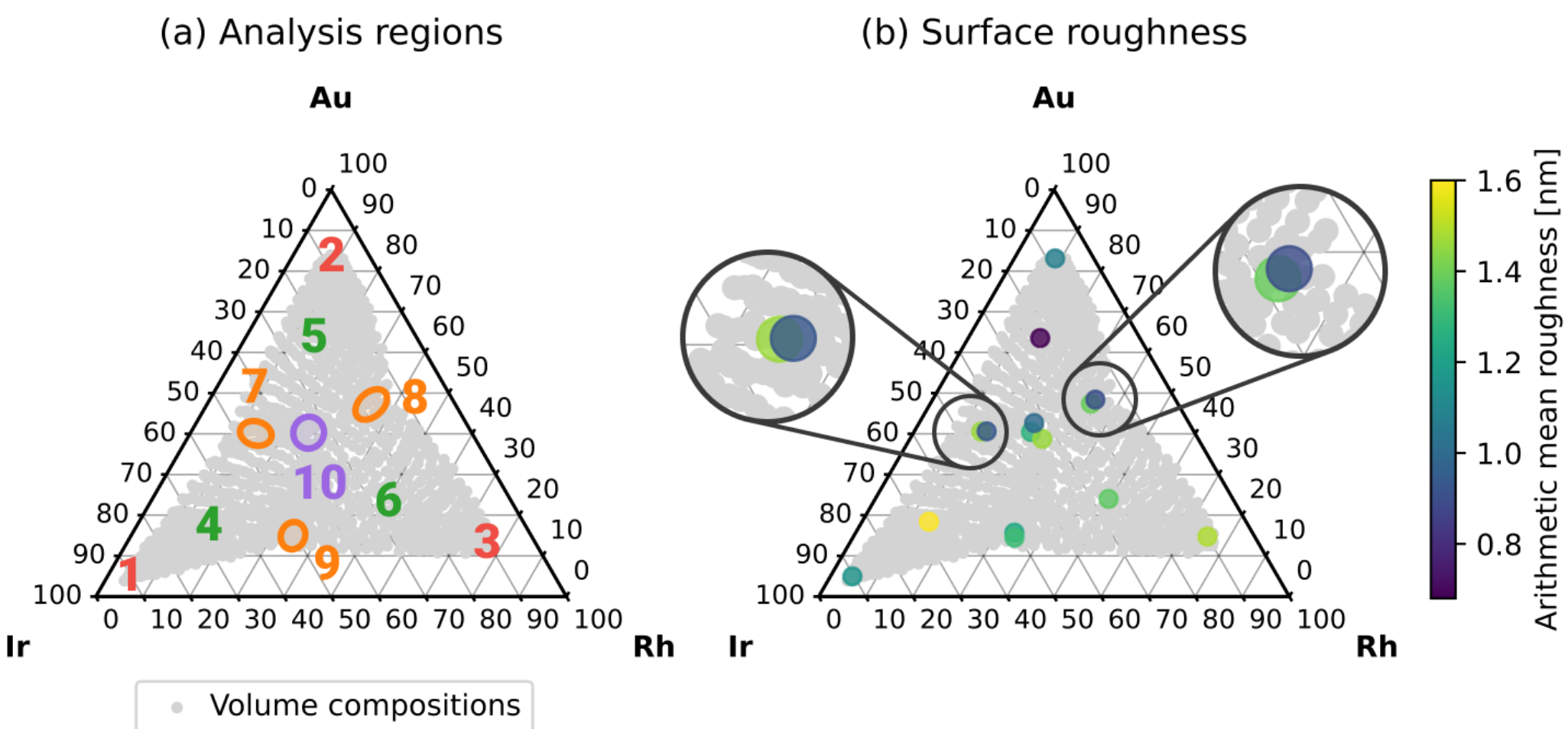

**Figure S16:** AFM-based surface roughness analysis across the Au–Ir–Rh composition space. (a) Overview of the 10 AFM analysis regions. Regions 1 (area 13 on the Ir-rich library), 2 (area 313 on Au-rich), and 3 (area 240 on Rh-rich) correspond to the compositionally extreme regions with the highest content of Au, Ir, and Rh. Regions 4–6 are located at the center of each individual library (measurement area 168 on all libraries). Regions 7–9 are situated in pairwise compositional overlap regions between libraries, where two AFM measurements were performed at each region (region 7: area 20 (Au-rich) and 315 (Ir-rich), region 8: area 178 (Au-rich) and 297 (Rh-rich), region 9: area 178 (Ir-rich) and 16 (Rh-rich)). Region 10 marks the near-equiatomic center of the ternary space, sampled on each of the three libraries (area 55 on the Au-rich library, 342 on the Ir-rich, and 200 on the Rh-rich one). (b) Arithmetic mean roughness $R_a$ (nm) of all 15 measurement areas, color-coded on the composition space. Regions 7 and 8 were magnified for increased visibility of the overlap.

The measured arithmetic mean roughness ranges from 0.7–1.6 nm across all 15 positions (Figure S16b), confirming that all films are smooth at the nanoscale and that the absolute difference in surface area between the compositions is small. Considering the distribution of the standard rate constant $k^0$ varying over multiple orders of magnitude over the explored composition space, the surface morphology differences do not represent a significant confounding factor in the interpretation of composition–activity trends as described in the main text. The roughness distribution also does not follow the smooth compositional gradients observed for $k^0$ and $\alpha$. The Au-rich library shows consistently lower roughness values (0.7–1.2 nm), notably also in the overlapping regions with the other libraries. In contrast, the Ir-rich and Rh-rich libraries show higher roughness values (> 1.2 nm) and show consistency in their overlapping compositional regions (regions 9 and 10).

This absence of a smooth, composition-dependent roughness trend across the full ternary space requires careful interpretation: with 15 measurement areas of 500 x 500 $nm^2$ each sampled across a total library area of 235 $cm^2$, the coverage is sparse, and the present measurements are not intended to provide a comprehensive mapping of surface morphology across the composition space. The inter-library roughness offset observed at the overlap regions nevertheless suggests that deposition parameters, and specifically the sputtering powers which were adjusted between library fabrication to achieve different composition spreads, may influence surface morphology independently of alloy composition.

AFM images recorded at all 15 measurement areas are compiled in Figures S17–S20, organized by analysis region.

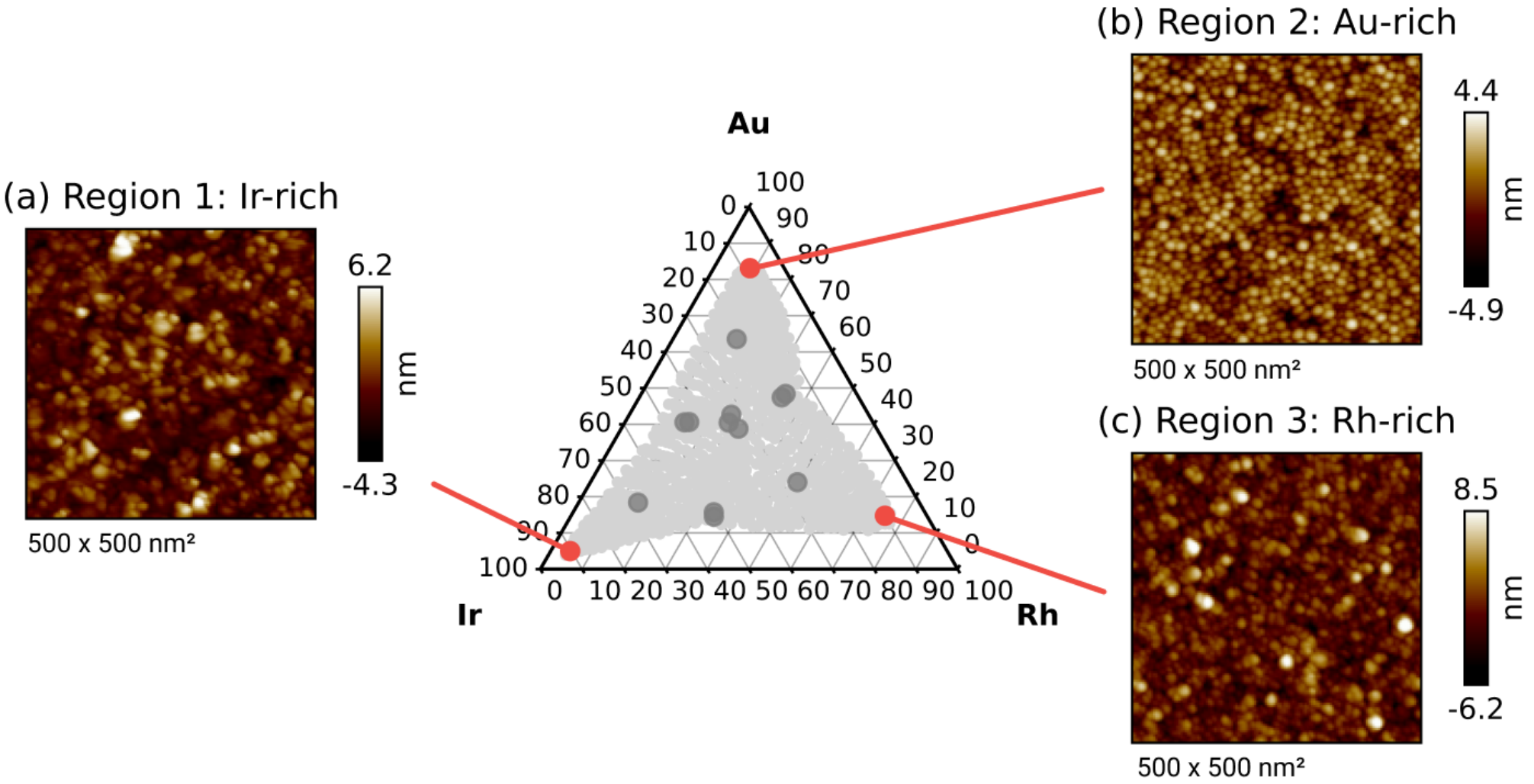

**Figure S17:** AFM images of compositions with highest content of each of the three elements (Regions 1-3). (a) Region 1, recorded at measurement area 13 of the Ir-rich library. (b) Region 2, recorded at area 313 of the Au-rich library. (c) Region 3, obtained from area 24 of the Rh-rich library.

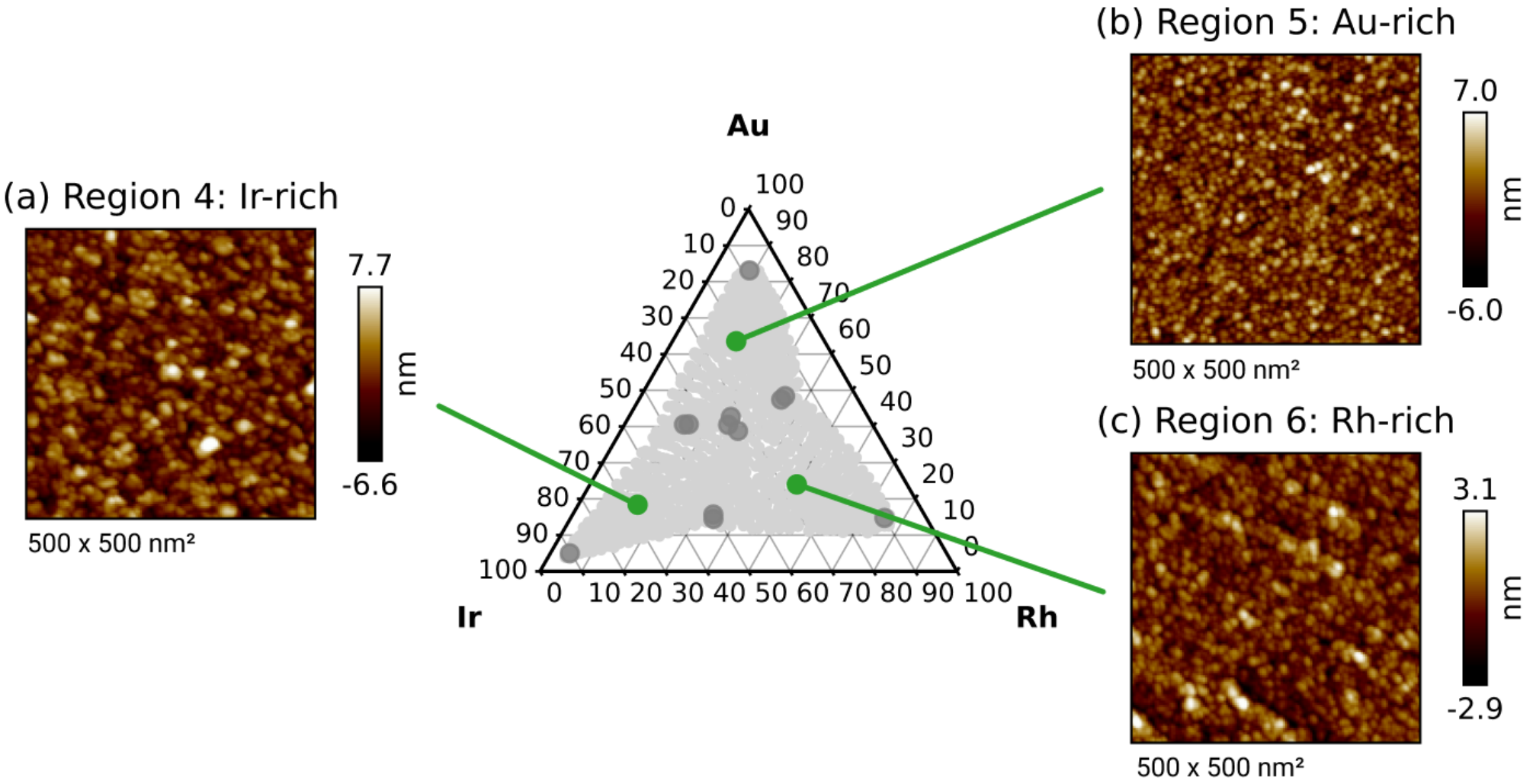

**Figure S18:** AFM images in the center of each of the three materials libraries (regions 4-6) at measurement area 168 of the (a) Ir-rich, (b) Au-rich, and (c) Rh-rich libraries.

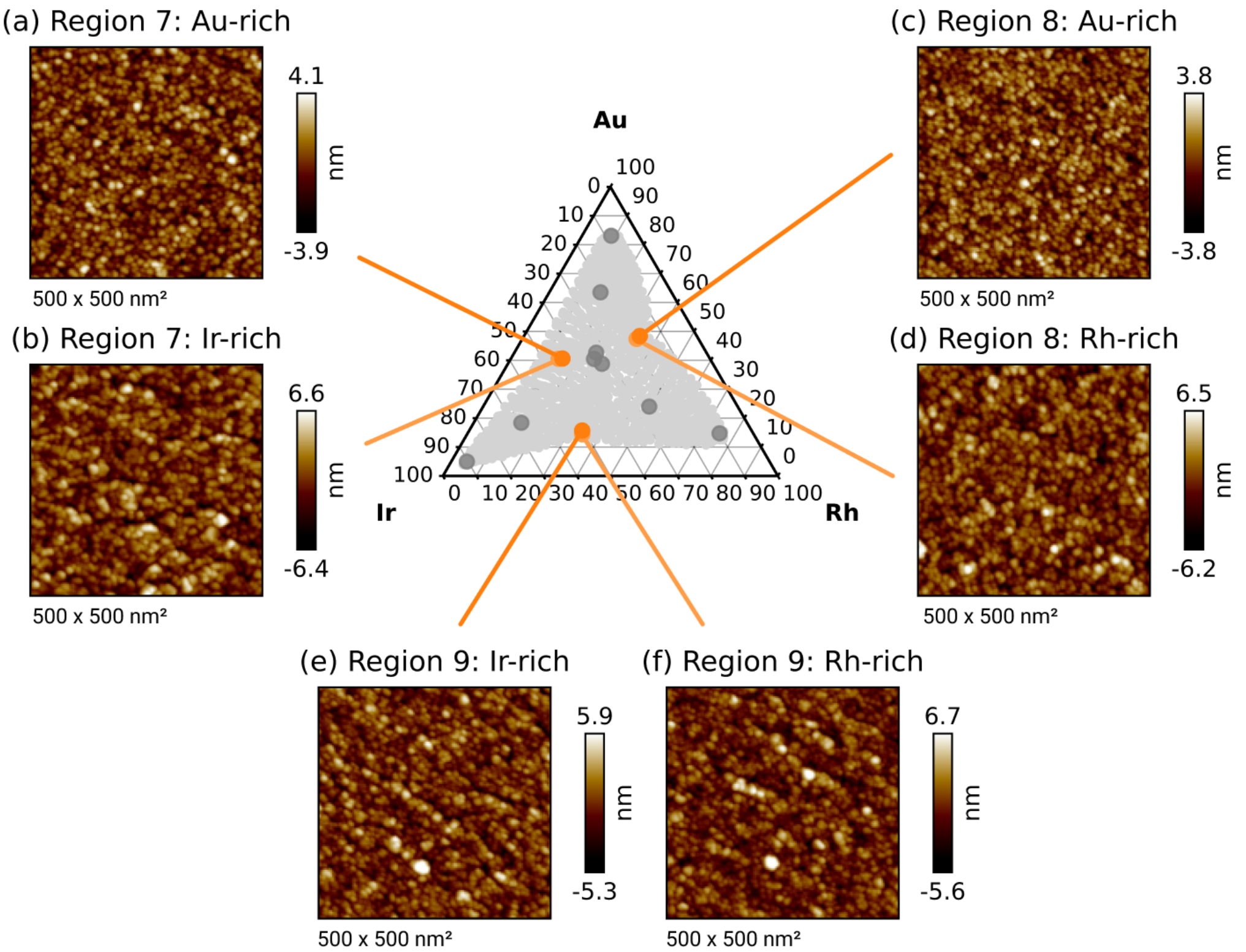

**Figure S19:** AFM images of pairwise compositional overlap regions (regions 7-9). (a,b) Region 7, recorded at measurement area 20 of the Au-rich library (a) and area 315 of the Ir-rich library (b). (c, d) Region 8, obtained from area 178 of the Au-rich library (c) and area 297 of the Rh-rich library (d). (e, f) Region 9, recorded at measurement area 178 of the Ir-rich library (e) and area 16 of the Rh-rich library (f). Each pair of images represents nominally similar compositions.

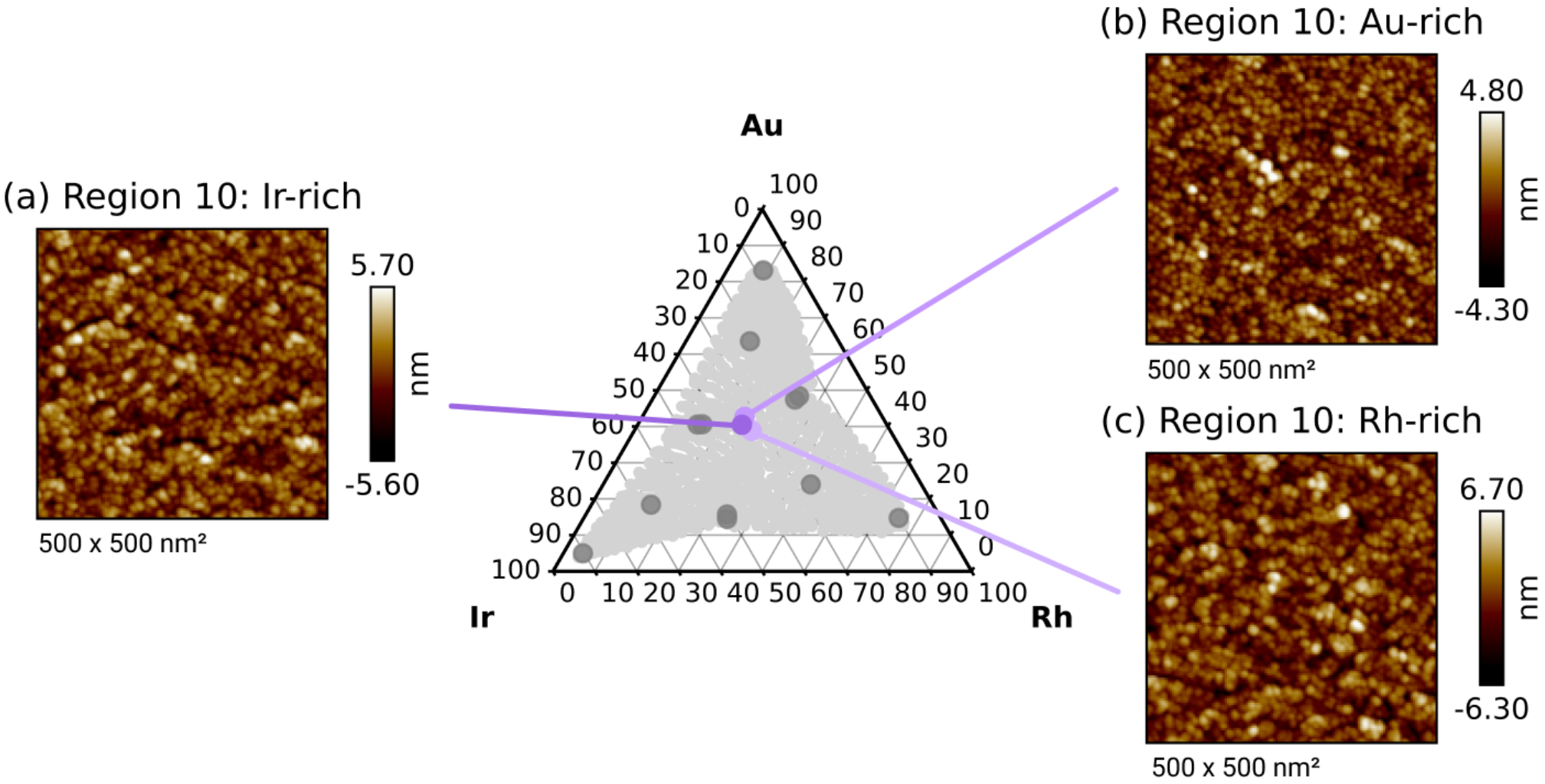

**Figure S20:** AFM images of the ternary composition space center (region 10), obtained from each of the three libraries: (a) measurement area 342 of the Ir-rich library, (b) area 55 of the Au-rich library, and (c) area 200 of the Rh-rich library.

## 2.5. Analysis of the performance of the autonomous SECCM measurement

Figure S21 shows the evolution of the limiting current over the measurement order, revealing a dependence on library transfers. During each transfer, the capillary is removed briefly (4 min) from the controlled Ar-atmosphere of the environmental chamber for repositioning and tilt correction. Partial drying of the electrolyte meniscus during this interval likely alters the effective wetted contact area before the next approach, thereby shifting the measured limiting current. Within each library, the limiting current shows high-frequency scatter reflecting the compositionally distributed, non-sequential sampling of the library due to the active learning algorithm. Superimposed with this is a low baseline drift, which could be associated with a possible progressive surface oxidation of the thin-film libraries over the duration of the measurement (26 h).

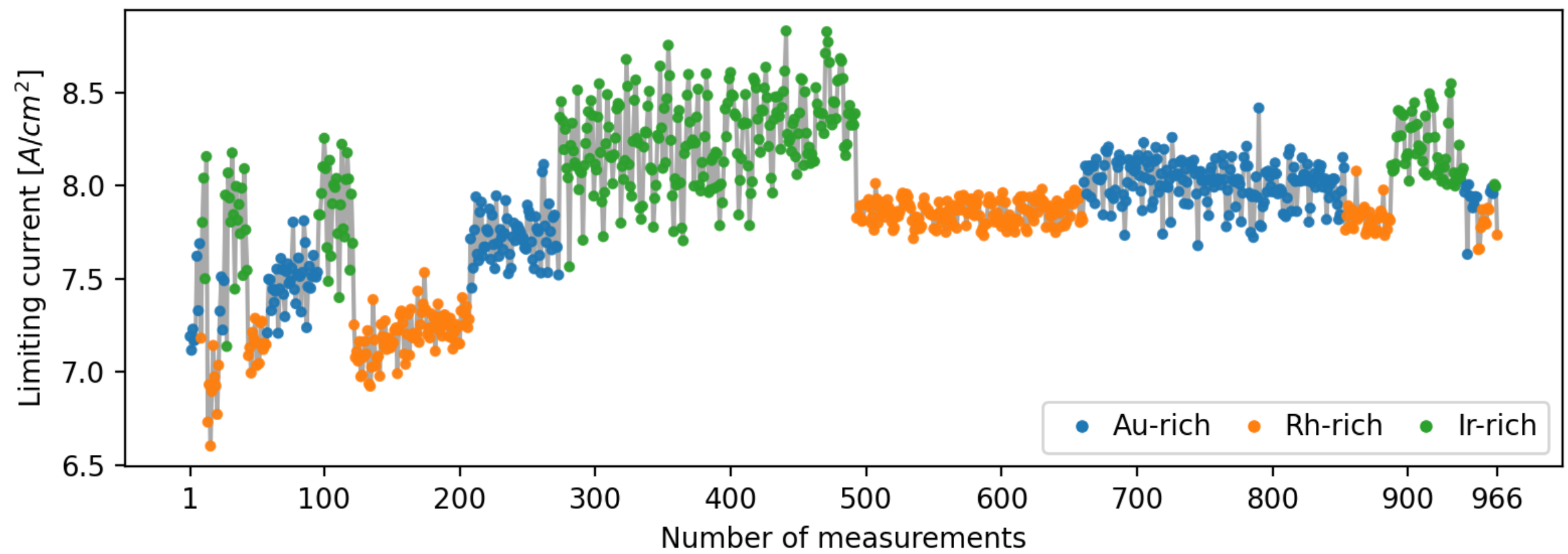

**Figure S21:** Limiting current density as a function of the measurement index across all 966 areas, colored by library. Vertical steps in the limiting current coincide with the transitions between libraries, reflecting brief exposure of the capillary to ambient conditions during the sample transfer. Within each library, the limiting current shows a high-frequency scatter arising from the non-sequential sampling pattern across different parts of the compositional gradient. In addition, there is a slow low amplitude drift likely due to continuing oxidation.

### 2.6. Original active learning of HER electrocatalytic activity in Au–Ir–Rh CCSS

The active learning performance is evaluated retrospectively against the ground truth of all 966 measurements and summarized in Figure S22. The prediction accuracy is quantified by the mean absolute error (MAE) of $k^0$ and $\alpha$, and normalized to the respective parameter ranges to allow direct comparison. The MAE of $k^0$ begins at 17% after the initialization measurements and rises slightly to 20% during the first few iterations, due to reflecting the initially limited compositional coverage of the initial dataset with respect to the entire space. It subsequently decreases rapidly, reaching 5% and plateauing shortly by iteration 100, before gradually approaching 1% over the remaining iterations. The MAE of $\alpha$ follows a different trajectory: it starts at 15% and drops sharply to 10% after the second iteration, before increasing again to 14% until iteration 100. From there, it decreases gradually and asymptotically approaches 5%.

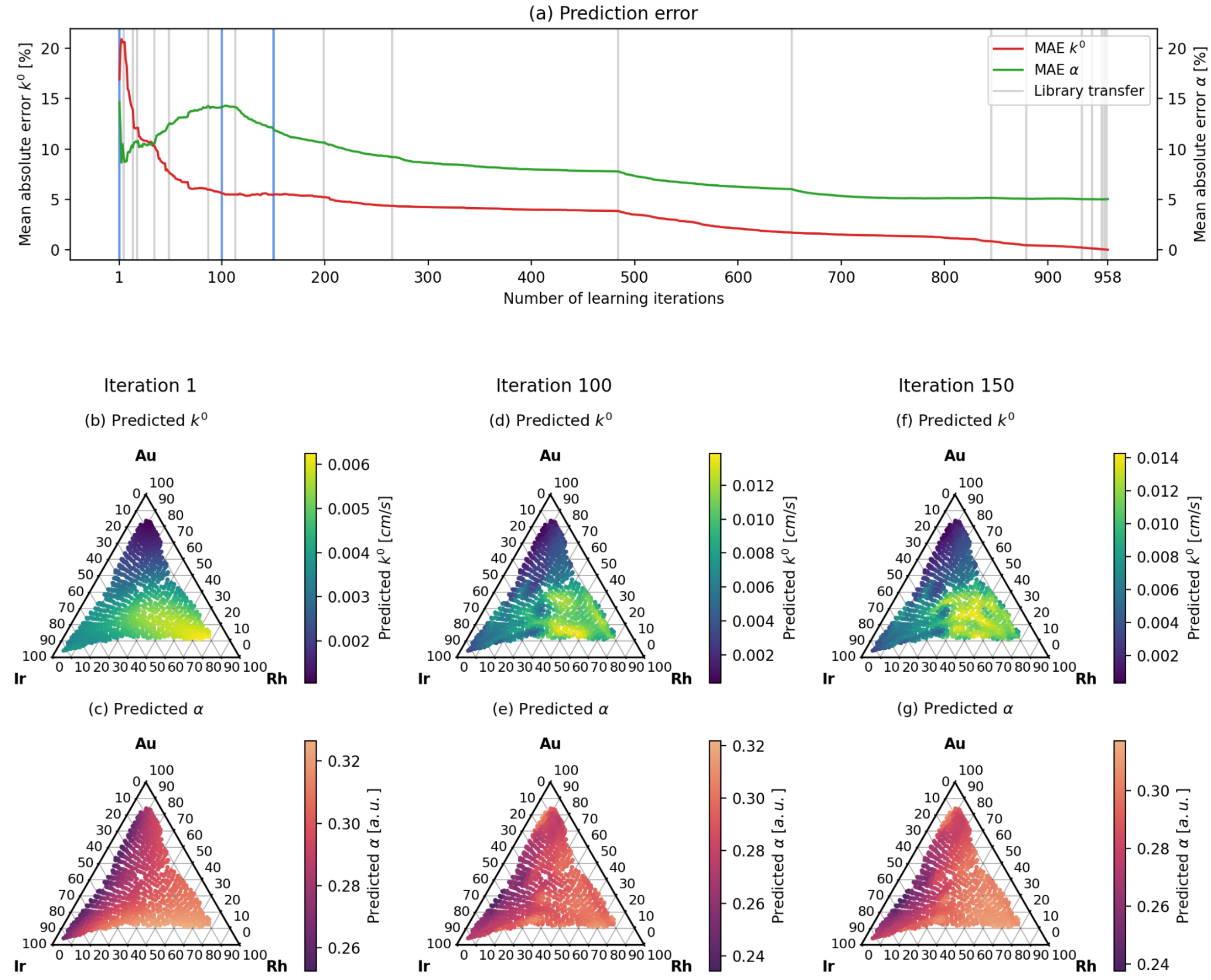

**Figure S22:** Active learning performance and evolution of the predictions across the Au–Ir–Rh composition space. (a) Mean absolute error (MAE) of the predicted $k^0$ and $\alpha$, each normalized to their respective parameter ranges ($k^0$: 0.0003–0.0142 cm/s, $\alpha$: 0.237–0.328) and expressed as a percentage, over the number of learning iterations. The MAE is evaluated against the ground truth of all 966 measurements. The vertical grey lines indicate library transfer events. (b) Predicted $k^0$ distribution in the ternary composition space after the first active learning iterations, when the model has been trained on the 8 initialization areas only. (c) Corresponding predicted $\alpha$ distribution after the first iteration. (d,e) Predicted distributions of $k^0$ and $\alpha$ after 100 iterations and (f,g) after 150 iterations.

The library transfer events, indicated in Figure S22a by vertical lines, occur more frequently at the beginning of the autonomous measurement sequence. Following the initialization measurements on the first (Ir-rich) library, the algorithm immediately transfers to the Rh-rich library, performs a single measurement, and then switches to the Au-rich library. This behavior is illustrated by Figure S22b, showing the measurement order on the composition space of the first 50 iterations. This rapid library switching arises from the epistemic uncertainty of the Gaussian process model being sufficiently large in the distant corners of the ternary composition space and therefore outweighing the acquisition cost penalty associated with the library transfer. The algorithm therefore prioritizes informational gain over measurement efficiency in the early stage, sacrificing some throughput to rapidly reduce the uncertainty across the full composition space. As measurements accumulate and the model uncertainty is reduced, the cost of library transfers begins to dominate the acquisition function, causing the device to remain on a single library for longer before switching.

The evolution of the model predictions is illustrated in Figures S22c-h, which show the predicted distributions of $k^0$ and $\alpha$ on the composition space at iterations 1, 100, and 150. Even after a single iteration, the predicted distributions of both $k^0$ and $\alpha$ already resemble the ground truth, correctly capturing the global trends. Over subsequent iterations, the model makes incremental adjustments, primarily refining local absolute values rather than altering relative trends. This behavior is consistent with the measurement-order-dependent drift in both parameters: as the model accumulates measurements across libraries and time, it progressively adapts its absolute predictions to account for this drift. The relative trends, which are of primary interest for combinatorial materials discovery, remain stable from early in the campaign.

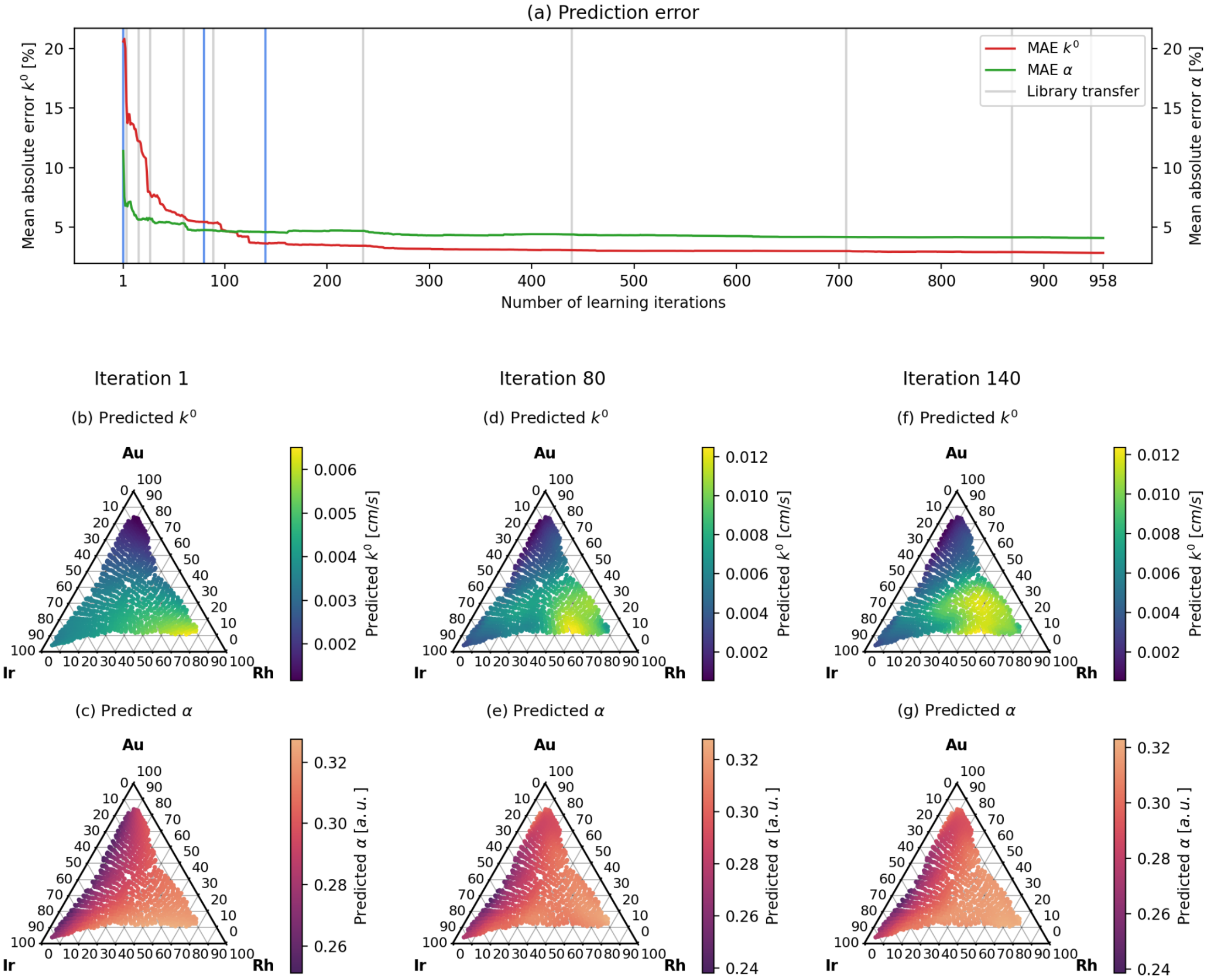

**Figure S23:** Active learning performance of the uncertainty-aware model and evolution of the predictions across the Au–Ir–Rh composition space. (a) Mean absolute error (MAE) of the predicted $k^0$ and $\alpha$, each normalized to their respective parameter ranges and expressed as a percentage, over the number of learning iterations. The MAE is evaluated against the ground truth of all 966 measurements. The vertical grey lines indicate library transfer events. (b) Predicted $k^0$ distribution in the ternary composition space after the first active learning iterations, when the model has been trained on the 8 initialization areas only. (c) Corresponding predicted $\alpha$ distribution after the first iteration. (d,e) Predicted distributions of $k^0$ and $\alpha$ after 80 iterations and (f,g) after 140 iterations. In comparison to the predictions of the original model, the distributions especially of $k^0$ and, but also of $\alpha$ are much smoother as the model can access the measurement uncertainty as additional information for the predictions. The uncertainty-aware model also no longer overfits, indicated by the MAE over the number of iterations not approaching zero when the model has consumed most of the training data.

Comparing the predictions of $k^0$ and $\alpha$ shows that the Gaussian process model appears to fit local variations more closely for $k^0$ than for $\alpha$. This is reflected in Figure S22a, where the MAE of $k^0$ approaches 0% while that of $\alpha$ stabilizes at ~5%, suggesting that the model overfits $k^0$. This behavior is a consequence of the logarithmic transformation applied to $k^0$. While beneficial to regress the exponential structure of the standard rate constant, the log transformation compresses the range of the data from nearly two orders of magnitude in linear space to a span of ~4 in log-space. This compression reduces the apparent replicate variance relative to the signal range, causing the noise parameter learned by the Gaussian process to converge to an artificially small value. This causes the model to interpret local fluctuations, which partly reflect measurement noise and drift rather than genuine compositional features, as true signal instead of smoothing them. A further contributing factor is the order in which averaging and log-transformation were applied in the original workflow: three LSVs were averaged per measurement area before extracting the LSV parameters and the log-transformation of $k^0$. While the mean is approximately preserved under this operation, the variance is not, since $var(log(k^0))$ cannot be recovered from $log(var(k^0))$. Therefore, the model misinterprets the noise variance, additionally contributing to the overfitting.

Figure S23 shows the active learning performance of the noise-aware Gaussian process model, which receives measurement variance of the three LSVs per measurement area as part of the Gaussian process likelihood. In contrast to Figure S22, the MAE of $k^0$ no longer approaches 0% and the predictions exhibit visibly smoother trends.

Since the ground truth is unavailable during an autonomous SECCM measurement, the MAE cannot be used as an indicator for a stopping criterion in regular campaigns. The mean predictive variance can serve as a real-time stopping criterion. In the present dataset, this mean predictive variance plateaus at approximately the same iteration count at which the MAE of $k^0$ converges. Figure S24 shows the predictive variance over the number of iterations. Both $k^0$ and the mean predicted variance plateau around 140 iterations, making it a suitable stopping criterion for this campaign. The authors note that a single test, as done here in case of the active learning campaign in the Au–Ir–Rh composition space, is not enough to fully establish a stopping criterion since tests must be performed on a wide range of materials systems to ensure appropriate stopping for future campaigns.

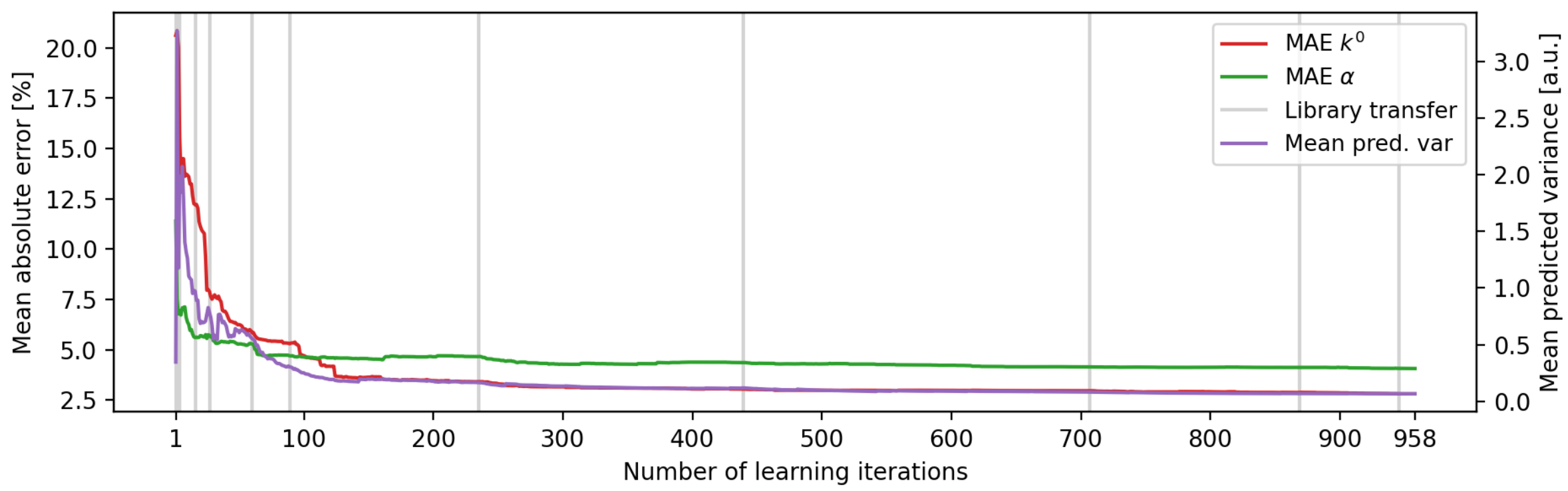

**Figure S24:** Mean predicted variance together with the prediction error of $k^0$ and $\alpha$. Both $k^0$ and the mean predicted variance plateau at around 140 iterations, making the mean predicted variance a suitable stopping criterion.

Figures S25 and S26 are available in ZENODO at https://doi.org/10.5281/zenodo.20439519 as animations. These show the active learning training data, predictions and acquisition function values during each training iteration both for the original Gaussian process model (Figure S25) as well as the noise-aware model (Figure S26).

## References

[1] E.B. Tetteh, O.A. Krysiak, A. Savan, M. Kim, R. Zerdoumi, T.D. Chung, A. Ludwig, W. Schuhmann, Long-Range SECCM Enables High-Throughput Electrochemical Screening of High Entropy Alloy Electrocatalysts at Up-To-Industrial Current Densities, Small Meth. 8 (2024) 2301284. https://doi.org/10.1002/smtd.202301284.

[2] S. Suzuki, K. Abe, Topological structural analysis of digitized binary images by border following, Comp. Vis., Graphics, Image Proc. 30 (1985) 32–46. https://doi.org/10.1016/0734-189X(85)90016-7.

[3] G. Arruda De Oliveira, M. Kim, C.S. Santos, N. Limani, T.D. Chung, E.B. Tetteh, W. Schuhmann, Controlling Surface Wetting in High-Alkaline Electrolytes for Single Facet Pt Oxygen Evolution Electrocatalytic Activity Mapping by Scanning Electrochemical Cell Microscopy, Chem. Sci. 15 (2024) 16331–16337. https://doi.org/10.1039/D4SC04407J.

[4] K.L. Anderson, M.A. Edwards, Evaluating Analytical Expressions for Scanning Electrochemical Cell Microscopy (SECCM), Anal. Chem. 95 (2023) 8258–8266. https://doi.org/10.1021/acs.analchem.3c00216.

[5] A.G. de G. Matthews, M. van der Wilk, T. Nickson, K. Fujii, A. Boukouvalas, P. León-Villagrá, Z. Ghahramani, J. Hensman, GPflow: A Gaussian Process Library using TensorFlow, J. Machine Learning Res. 18 (2017) 1–6.

[6] C.E. Rasmussen, C.K.I. Williams, Gaussian Processes for Machine Learning, The MIT Press, 2005. https://doi.org/10.7551/mitpress/3206.001.0001.

[7] K.P. Murphy, Machine Learning: A Probabilistic Perspective, MIT Press, Cambridge, MA, USA, 2012.

[8] F. Thelen, F. Lourens, A. Ludwig, Accelerating Surface Composition Characterization of Thin-Film Materials Libraries Using Multi-Output Gaussian Process Regression, Adv. Intell. Discov. 2 (2026) e202500062. https://doi.org/10.1002/aidi.202500062.

[9] F. Thelen, R. Zehl, R. Zerdoumi, J.L. Bürgel, L. Banko, W. Schuhmann, A. Ludwig, Accelerating Combinatorial Electrocatalyst Discovery with Bayesian Optimization: a Case Study in the Quaternary System Ni-Pd-Pt-Ru for the Oxygen Evolution Reaction, Adv. Sci. 12 (2025) e07302. https://doi.org/10.1002/advs.202507302.

[10] E. Bonilla, K. Chai, C. Williams, Multi-task Gaussian Process Prediction, in: Advances in Neural Information Processing Systems, Curran Associates, Inc., 2007. https://papers.nips.cc/paper_files/paper/2007/hash/66368270ffd51418ec58bd793f2d9b1b-Abstract.html (accessed May 29, 2026).

[11] M. Greenacre, Compositional Data Analysis, Annu. Rev. Stat. Appl. 8 (2021) 271–299. https://doi.org/10.1146/annurev-statistics-042720-124436.

[12] J.J. Egozcue, V. Pawlowsky-Glahn, G. Mateu-Figueras, C. Barceló-Vidal, Isometric Logratio Transformations for Compositional Data Analysis, Math. Geol. 35 (2003) 279–300. https://doi.org/10.1023/A:1023818214614.

[13] N. Pukhareva, M. Kim, F. Thelen, G. Arruda De Oliveira, R. Zehl, W. Schuhmann, A. Ludwig, Compositional and Structural Impact on the Hydrogen Evolution Reaction Activity across Noble-Metal-Based Compositionally Complex Solid Solutions Thin Film

Libraries, ACS Electrochem. 2 (2026) 364–373. https://doi.org/10.1021/acselectrochem.5c00406.

[14] C.M. Clausen, T.A.A. Batchelor, J.K. Pedersen, J. Rossmeisl, What Atomic Positions Determines Reactivity of a Surface? Long-Range, Directional Ligand Effects in Metallic Alloys, Adv. Sci. 8 (2021) 2003357. https://doi.org/10.1002/advs.202003357.